\documentclass[a4paper,11pt]{article}

\usepackage{jheppub}

\usepackage[utf8]{inputenc}
\usepackage[T1]{fontenc}

\usepackage{amsmath}
\usepackage{amsfonts}
\usepackage{dsfont}
\usepackage{mathrsfs}
\usepackage{listings}
\usepackage{hyperref}
\usepackage{graphicx}
\usepackage{cleveref}
\usepackage{tensor}
\usepackage{todonotes}
\usepackage{braket}
\usepackage{mleftright} \mleftright

\newcommand{\eq}[2]{\begin{equation} #1 \label{#2} \end{equation}}

\newcommand*{\dif}{\mathrm{d}}

\newcommand*{\R}{\mathbb{R}}

\DeclareMathOperator{\tr}{tr}
\renewcommand{\L}{\mathcal{L}}
\newcommand*{\J}{\mathcal{J}}

\newcommand*{\G}{\mathcal{G}}
\newcommand*{\ve}{\varepsilon}
\newcommand*{\veb}{\bar{\varepsilon}}
\newcommand*{\vph}{\varphi}
\newcommand*{\bt}{\widetilde\beta}

\newcommand*{\Lt}{\mathtt{L}}
\newcommand*{\Wt}{\mathtt{W}}
\newcommand*{\Gt}{\mathtt{G}}
\newcommand*{\Jt}{\mathtt{J}}

\title{Null warped AdS in higher spin gravity}

\author[a,b]{Veronika Breunhölder,}
\author[a]{Mirah Gary,}
\author[a]{Daniel Grumiller}
\author[a]{and Stefan Prohazka}

\affiliation[a]{Institute for Theoretical Physics\\
          TU Wien\\
          Wiedner Hauptstrasse 8--10/136\\
          A-1040 Vienna, Austria}

\affiliation[b]{School of Mathematics and Maxwell Institute for Mathematical Sciences\\
University of Edinburgh\\
King’s Buildings\\
Edinburgh EH9 3JZ, U.K.}

\emailAdd{v.breunhoelder@ed.ac.uk}
\emailAdd{mgary@hep.itp.tuwien.ac.at}
\emailAdd{grumil@hep.itp.tuwien.ac.at}
\emailAdd{prohazka@hep.itp.tuwien.ac.at}

\abstract{
We equip three-dimensional spin-3 gravity in the principal embedding with a new set of boundary conditions that we call ``asymptotically null warped AdS''. We find a chiral copy of the Polyakov--Bershadsky algebra as asymptotic symmetry algebra, reminiscent of the situation in topologically massive gravity with strict null warped AdS boundary conditions. We prove the invertibility of the map between zuvielbein and metric variables and construct a global gauge transformation to half of AdS spin-3 gravity in the diagonal embedding. This explains why the theory is chiral and why the Polyakov--Bershadsky algebra arises. We then introduce chemical potentials, derive the entropy, free energy, and the holographic response functions, and conclude with a discussion.
}

\keywords{gauge-gravity correspondence, null warped AdS, higher spin gravity, three-dimensional gravity, Chern--Simons gauge theories}

\preprint{TUW--15--19}

\begin{document}

\maketitle
\flushbottom

\section{Introduction}
\label{sec:introduction}

Many current research directions address the question of how general holography is \cite{BBB}. 

The holographic principle \cite{'tHooft:1993gx,Susskind:1994vu}, originally motivated by the extensive behavior of black hole entropy in one dimension lower than expected from a quantum field theory perspective, has found a concrete realization in the Anti-de~Sitter/Conformal Field Theory (AdS/CFT) correspondence \cite{Maldacena:1997re,Gubser:1998bc,Witten:1998qj,Aharony:1999ti}. It is interesting to ponder whether the key insights about holography so far are specific to AdS/CFT or if they are general lessons for quantum gravity. In order to address this issue one needs to go beyond the usual AdS/CFT correspondence. 

An example for a generalization in this spirit is to consider non-unitary AdS/CFT, in order to find out if holography can work also for non-unitary theories, see \cite{Grumiller:2008qz,Grumiller:2013at,Vafa:2014iua} for some proposals. Another example is to consider (again in an AdS/CFT context) Vasiliev-type higher-spin holography \cite{Sezgin:2002rt,Klebanov:2002ja}, particularly in three dimensions \cite{Henneaux:2010xg,Campoleoni:2010zq,Gaberdiel:2010pz}. An advantage of three-dimensional higher-spin gravity models is that the massless higher spin fields share a useful property with their spin-2 cousin: they are all pure gauge locally. In this sense, higher spin fields are not more exotic than spin-2 fields in three dimensions. 

In this paper we focus on three-dimensional higher spin gravity. However, before presenting our model and the boundary conditions that are part of its definition we turn to a wider class of generalizations of AdS/CFT, in order to motivate our specific choices.

Dropping the assumption that spacetime is asymptotically AdS leads to a variety of additional possibilities, including flat space holography \cite{Susskind:1998vk,Polchinski:1999ry,Giddings:1999jq,Gary:2009mi}, de~Sitter holography \cite{Strominger:2001pn,Anninos:2011ui}, Lifshitz holography \cite{Kachru:2008yh} and Schrödinger holography \cite{Son:2008ye,Balasubramanian:2008dm}. If holography is a true aspect of nature it should work beyond AdS/CFT. Thus, the study of holography in these spacetimes becomes pertinent to the question in the first paragraph.

Besides this strong theoretical motivation to study non-AdS holography, there are also good phenomenological reasons to go beyond AdS/CFT. Namely, several condensed matter systems show anisotropic scale invariance near some fixed point, so that they cannot be described by usual relativistic CFTs (see \cite{sachdev2011quantum} and Refs.~therein). If one attempts to construct holographic duals for these field theories (to describe certain strong coupling phenomena and transport properties) the bulk spacetime cannot asymptote to AdS, which then clearly requires non-AdS holography. Indeed, this line of reasoning was the main motivation for seminal work on Lifshitz and Schrödinger holography \cite{Kachru:2008yh,Son:2008ye,Balasubramanian:2008dm}.

Some holographic correspondences can be set up in a puristic way, by which we mean that essentially only the metric is needed as a field to achieve the desired asymptotic behavior of spacetime. This includes, in addition to AdS/CFT, flat space holography and dS/CFT. By contrast, Lifshitz and Schrödinger holography typically require the introduction of (exotic) matter fields in addition to the metric or some higher derivative interactions, since pure Einstein gravity does not support such spacetimes.

A few years ago two of the present authors realized (together with Radoslav Rashkov) that higher-spin theories could provide an avenue for non-AdS holography without the introduction of any additional matter fields or higher derivative interactions \cite{Gary:2012ms}. The three-dimensional higher-spin theories considered in that paper support solutions that can have a variety of asymptotic behaviors, including Lifshitz and Schrödinger. Thus, attempts to set up certain types of non-AdS holography naturally led to higher-spin gravity in three dimensions.

One of the many questions that was left unanswered in the original work \cite{Gary:2012ms} was the existence of suitable boundary conditions that allow for the relevant non-AdS solutions. Several papers in the past few years studied this question (and related ones), see e.g.~\cite{Afshar:2012hc,Gutperle:2013oxa,Afshar:2014cma,Gutperle:2014aja,Beccaria:2015iwa,Lei:2015ika,Lei:2015gza}, and some were indeed able to find consistent sets of boundary conditions that permit these solutions, including Lobachevsky holography \cite{Afshar:2012nk}, flat space holography \cite{Afshar:2013vka,Gonzalez:2013oaa} and Lifshitz holography \cite{Gary:2014mca}.

However, there are some conceptual subtleties in all these constructions. First of all, the metric is not gauge invariant in higher-spin theories, so labeling some set of configurations as ``asymptotically something'' can be misleading, whatever ``something'' might be. The solution to this issue would be a generalization of Riemann calculus that allows an invariant characterization of all relevant aspects of higher spin geometries, but this problem has not been solved yet. Thus, the best one can do at the moment is to avoid the metric formulation as much as possible and to phrase everything in terms of zuvielbein variables (and gauge invariant combinations thereof). See~\cite{Campoleoni:2014tfa} and Refs.~therein for a summary on the metric approach in three-dimensional spin-3 gravity.

This leads to the second potential subtlety: even though we have to accept that the metric and its asymptotic behavior are gauge dependent, for several purposes we would like to have at least some metric interpretation available. However, the map between zuvielbein and metric variables can fail to be invertible. Indeed, as shown by Lei and Ross \cite{Lei:2015ika}, this situation does actually arise for the Lifshitz configurations introduced in \cite{Gary:2014mca}. It should be stressed that these configurations are perfectly well-behaved and regular, but the interpretation as ``asymptotically something'' becomes even less clear if there is no invertible map between first order (zuvielbein) and second order (metric) variables. 

The third potential subtlety concerns the symmetries of the field theory. Even if we grant that the gravity side may not allow for a unique label like ``AdS holography'', ``Lifshitz holography'' or something else, the field theory side should give a decisive clue how to appropriately label the holographic correspondence, since the symmetries are uniquely defined on both sides of the correspondence. For instance, encountering anisotropic scale invariance would allow justifiably attaching the label ``Lifshitz'' or ``Schrödinger''. However, as was shown by three of the present authors (together with Soo-Jong Rey) in higher-spin theories anisotropic scale invariance can get extended to an isotropic one \cite{Gary:2014mca}, which again complicates the interpretation.

Since the three subtleties above are mostly about interpretation and not about internal consistency of the theory, its boundary conditions and associated asymptotic symmetries, one may disregard them as idle. But when trying to address interesting questions such as ``does holography also work in non-AdS spacetimes?'' it is at least important to discriminate between AdS and non-AdS. Therefore, it is of relevance to confront these issues.

In this paper we construct explicitly a new example that touches all the subtleties above and differs qualitatively from all constructions so far. Namely, we consider three-dimensional spin-3 gravity in the principal embedding with boundary conditions that one may suggestively call ``null warped AdS boundary conditions'', since the metric asymptotes to null warped AdS$_3$ (see \cite{Detournay:2005fz,Anninos:2008fx,Jeong:2014iva} for some previous constructions with massive gravitons or exotic matter). We study this model in detail and address the issues raised above.

This paper is organized as follows.
In section \ref{sec:null-warped-bcs} we review some aspects of null warped AdS geometries and spin-3 gravity, and then define our theory and its boundary conditions.
In section \ref{sec:consistency} we show the consistency of our theory by constructing the canonical charges and their symmetry algebra, and by proving the invertibility of the map between first and second order variables. Moreover, we construct a global gauge transformation to a chiral half of an asymptotically AdS configuration.
In section \ref{sec:thermo} we discuss the thermodynamics, starting by the addition of chemical potentials, and then calculating entropy, which we bring into chiral Cardy form. We also study the first law, free energy and holographic response functions.
In section \ref{sec:discussion} we conclude with a discussion of our results.

%==================================================

%==================================================

\section{Definition of the theory}
\label{sec:null-warped-bcs}

In this section we start by reviewing essential properties of null warped AdS geometries in section \ref{sec:null-warped-ads}, continue with relevant aspects of the Chern--Simons formulation of higher spin gravity in section \ref{sec:CS} and then formulate our boundary conditions in section \ref{sec:bcs}. In section \ref{sec:bcpt} we display all boundary condition preserving transformations for later use.

\subsection{Null warped AdS geometry}
\label{sec:null-warped-ads}

In the present work the geometries of interest are null warped AdS. They are a special case of a larger class of geometries studied by a number of researchers mainly in the context of topologically massive gravity \cite{Deser:1982vy,Deser:1981wh,Deser:1982a}, see e.g.~\cite{Clement:1994sb,Deser:2004wd,Detournay:2005fz,Anninos:2008fx,Gibbons:2008vi,Ertl:2010dh,Anninos:2010pm}:
\begin{equation}
  \label{eq:nullmetric}
  \dif s^{2}_{\mathrm{(z)}}/\ell^{2}=\frac{\dif r^{2}}{4 r^{2}} +2 r \,\dif t\, \dif \varphi+ f(r,\,z)\, \dif \varphi^{2} 
\end{equation}
with curvature radius $\ell$, angular coordinate $\varphi \sim \varphi+2 \pi$, time coordinate $t\in (-\infty,\infty)$ and radial coordinate $r\in(r_{s},\infty)$. The quantity $r_s$ is the largest negative real root of the function $f$, which depends on the radial coordinate $r$ and a real parameter $z$,
\begin{equation}
  \label{eq:f}
  f(r,\,z)=r^{z}+\beta r + \alpha^{2}
\end{equation}
as well as on the constants of motion $\alpha$ and $\beta$. All the spacetimes \eqref{eq:nullmetric} have a Killing horizon at $r=0$, while the asymptotic region corresponds to the limit $r\to\infty$. For a comprehensive discussion of these geometries and holographic implications within topologically massive gravity see \cite{Anninos:2010pm}.

In topologically massive gravity the parameter $z$ depends on the coupling constant of the gravitational Chern--Simons term. Asymptotically null warped AdS is the special case $z=2$, on which we shall focus from now on. In that case the coordinate transformation $r=e^{2 \rho}$ with a rescaling of $t$ by a factor $\tfrac12$ brings the line-element \eqref{eq:nullmetric} into the useful form
\begin{equation}
  \label{eq:nullusef}
    \dif s^2/\ell^{2}=\dif \rho^{2} + e^{2 \rho}\, \dif t\, \dif \varphi+ \big(e^{4 \rho}+ \beta e^{2 \rho} + \alpha^{2} \big)\,\dif \varphi^{2} \,.
\end{equation}
Depending on the parameters $\alpha$ and $\beta$ the geometry defined by the asymptotically null warped AdS line-element \eqref{eq:nullusef} belongs to one of the following classes:
\begin{itemize}
 \item Null warped AdS vacuum: $\alpha=0=\beta$
 \item Null warped AdS black holes: $\alpha^2\geq 0$, $\beta\geq 2|\alpha|$
 \item Null warped AdS solitons: $\alpha^2>\beta^2/4$, $\beta\geq 0$
 \item Null warped AdS naked singularities: $\beta<0$ or $\alpha^2<0$
\end{itemize}
The naked singularities are closed time-like curves, which arise because for $\beta<0$ the largest root of $f=0$ has a positive value, $r_s>0$, and thus is not screened by the Killing horizon at $r=0$ (similar remarks apply to $\alpha^2<0$). The main difference between black holes and solitons is that the former have closed time-like curves hidden behind a horizon, while the latter have no closed time-like curves.

\subsection{Chern--Simons formulation of spin-3 gravity}\label{sec:CS}

For a large class of $(2+1)$-dimensional spin-3 gravity theories it
is possible to write the respective action
\cite{Blencowe:1988gj,Hoppephdthesis,Bergshoeff:1989ns,Bordemann:1989zi,Henneaux:2010xg,Campoleoni:2010zq}
as
\begin{equation}
  I = I_{\text{CS}}[A] - I_{\text{CS}}[\bar{A}]+B
\label{eq:action}
\end{equation}
with Chern-Simons action
\begin{equation}
I_{\text{CS}}[A] = \frac{k}{4\pi} \int_{\mathcal{M}} \tr \Big(A \wedge \dif A +
  \frac{2}{3} A\wedge A\wedge A \Big)
\end{equation}
where $k=\tfrac{\ell}{16 G}$ is the Chern-Simons level, and a boundary term $B$ determined in the next section. The connections $A$ and $\bar{A}$ are commuting $sl(3,\R)$ valued one-forms, see Appendix \ref{sec:principal-embedding} for our conventions concerning the principal embedding of $sl(2)$ into $sl(3)$. The gauge connections $A$, $\bar A$ are related to the zuvielbein $e$ and the dualized spin connection $\omega$ via
\begin{equation}
  A_\mu = \omega_\mu + \frac{1}{\ell} e_\mu \qquad \bar{A}_\mu =
  \omega_\mu - \frac{1}{\ell} e_\mu
\end{equation}
where $\ell$ is the curvature radius.
The equations of motion (EOM) are given by gauge flatness conditions for the
connections
\begin{equation}
\label{eq:EOM}
  F = \dif A + A\wedge A = 0 = \dif \bar{A} + \bar{A}
  \wedge \bar{A} =\bar{F}\,.
\end{equation}

\subsection{Null warped AdS boundary conditions in spin-3 gravity}\label{sec:bcs}

By a partial gauge fixing, $A$ and $\bar{A}$ can be
written as \cite{Banados:1994tn, Campoleoni:2010zq}
\begin{equation}
  A = b^{-1} \dif b + b^{-1} (\hat{a}^{(0)} + a^{(0)} +
  a^{(1)} ) b, \qquad
  \bar{A} = b \dif b^{-1} + b (\hat{\bar{a}}^{(0)} +
  \bar{a}^{(0)} + \bar{a}^{(1)} ) b^{-1}
\label{eq:nolabel}
\end{equation}
where we chose $b = e^{\rho \Lt_0}$, such that
$A$ is decomposed into a fixed background $\hat{a}^{(0)}$, leading
order fluctuations $a^{(0)}$ and subleading terms $a^{(1)}$ (and
similarly for $\bar{A}$).

To construct null warped AdS we follow \cite{Gary:2012ms} and take as background connection
\begin{equation}
  \hat{a}^{(0)} = (\Lt_1 -\tfrac{2}{3} \Wt_2) \dif\vph, \qquad
  \hat{\bar{a}}^{(0)} = \tfrac{2}{3}\Wt_{-2} \dif\vph + \Lt_{-1} \dif t.
\end{equation}
The metric and the spin-3 field are defined in the standard way as (in
the following we set $\ell =1$)
\begin{align}
  \label{eq:gandphi}
  g_{\mu\nu} = \frac{1}{2} \tr (e_\mu e_\nu), \qquad  \phi_{\mu \nu \xi}=\frac{1}{3 !} \tr (e_{( \mu}e_{\nu} e_{\xi )} )
\end{align}
and lead to
\begin{align}
  \label{eq:nw02}
  \dif s^2 &=  \dif \rho^2 +  e^{2\rho} \,\dif t\, \dif \vph + 
  e^{4\rho}\, \dif \vph^2 \\
  \phi&= -\frac{3}{2} e^{4 \rho}
  \left(
    \dif t^{2} \dif \varphi+ \dif \varphi^{3}
  \right)
%  \phi_{tt \varphi}&=\phi_{t \varphi t }=\phi_{\varphi\varphi t}-\frac{1}{2} e^{4 \rho}   \quad \phi_{\varphi \varphi \varphi}=-\frac{3}{2} e^{4 \rho}.
\end{align}
The line element \eqref{eq:nw02} is the null warped AdS vacuum, see section \ref{sec:null-warped-ads} and compare with \eqref{eq:nullusef} for $\alpha=0=\beta$. The asymptotic EOM determine the general form of the state-dependent fluctuations in the $a_{\varphi}$ component of 
$a^{(0)}$ and $\bar{a}^{(0)}$ as
\begin{subequations}\label{eq:nw04}
\begin{align}
  a^{(0)} &= \big(l_{-1}(\vph)\Lt_{-1} + \widetilde{w}_{1}(\varphi) \Wt_{1}+ w_0(\vph) \Wt_0 +
            w_{-1}(\vph) \Wt_{-1}+ w_{-2}(\vph) \Wt_{-2}  \big)\dif\vph\\
  \bar{a}^{(0)} &= \widetilde{\beta}(\varphi) \Lt_{-1} \dif \varphi
\end{align}
\end{subequations}
leading to a line element
\begin{equation}
  \dif s^2 =  \dif \rho^2 +  e^{2\rho}\, \dif t\, \dif \vph + \Big(
    e^{4 \rho }+e^{2 \rho}\widetilde{\beta}- l_{-1}+\tfrac{3}{16} \big(w_0^2-8
      w_{-2}-3 \widetilde{w}_{1} w_{-1}\big)\Big)\,\dif \vph^2\,.
\label{eq:ds}
\end{equation}

The canonical analysis performed in the next section reveals that $\widetilde\beta$ and $\widetilde w_{1}$ are pure gauge. We show this for $\widetilde\beta$ explicitly below whereas, to reduce clutter, we set $\widetilde w_{1}$ to zero from the start. Then the line-element \eqref{eq:ds} simplifies to asymptotically null warped AdS \eqref{eq:nullusef} with a $\vph$-dependent function $\alpha$ and vanishing $\beta$,
\eq{
\alpha^2=-l_{-1}+\tfrac{3}{16} (w_0^2-8w_{-2})\qquad \beta=0\,. 
}{eq:angelinajolie}
Comparison with the discussion in section \ref{sec:null-warped-ads} shows that null warped AdS black holes are gauge equivalent to null warped AdS solitons and to null warped naked singularities in our theory. It is a well-known property of higher-spin theories that geometrically distinct configurations can be gauge equivalent, which is a key aspect of singularity resolutions. 

\subsection{Boundary condition preserving transformations and chirality}\label{sec:bcpt}

The boundary conditions \eqref{eq:nw04} are preserved by transformations that satisfy
\begin{align}
\label{eq:nw03}
  \delta A &= \dif \ve + [A , \ve] = \mathcal{O}(A) &   \delta \bar A &= \dif \veb + [\bar A , \veb] = \mathcal{O}(\bar A)
\end{align}
where, similar to the connection, the parameter $\ve$ and $\veb$ can be decomposed as
\begin{align}
  \label{eq:2}
    \ve&= b^{-1} \big(\ve^{(0)} + \ve^{(1)}\big) b, & \veb &= b \big(\veb^{(0)} + \veb^{(1)}\big) b^{-1}\,.
\end{align}
Ignoring any subleading contributions, by solving \eqref{eq:nw03} one
finds the most general boundary condition preserving 
transformation  for the non-barred sector are generated by a parameter of the
form
\begin{multline}\label{eq:nw01}
  \ve^{(0)} = \ve_{\Lt_1}\Lt_1 
  + \frac{3}{4}\left(\ve_{\Wt_2}' + \ve_{\Wt_1}\right)\Lt_0 
  + \frac{3}{8}\left( \ve_{\Wt_1}' -2w_0\ve_{\Lt_1} - 4l_{-1}\ve_{\Wt_2}+ 2\ve_{\Wt_0}\right)\Lt_{-1}  \\
  + \ve_{\Wt_2}\Wt_2 
  + \ve_{\Wt_1}\Wt_1 
  + \ve_{\Wt_0}\Wt_0 
  - \frac{1}{12}\left( 8\ve_{\Lt_1}'+6\ve_{\Wt_2}' +18w_{-1}\ve_{\Wt_2}+ (6-9w_0)\ve_{\Wt_1} \right) \Wt_{-1} \\
  -\frac{1}{4}\left(\tfrac{1}{2} \ve_{\Wt_2}'' + \ve_{\Wt_1}' - ( \tfrac{4}{3} l_{-1}+ w_0 ) \ve_{\Lt_1} - (2 l_{-1} - 6 w_{-2})\ve_{\Wt_2} - \tfrac{3}{4}w_{-1}\ve_{\Wt_1}+ \ve_{\Wt_0}   \right) \Wt_{-2}
\end{multline}
where $\ve_{\Lt_1}$, $\ve_{\Wt_2}$, $\ve_{\Wt_1}$ and $\ve_{\Wt_0}$ are
arbitrary functions of $\vph$. These functions multiply non-negative weight generators. 

We now proceed with the analysis for the barred sector, proving in particular also our earlier statement that $\widetilde \beta$ is pure gauge. Solving conditions \eqref{eq:nw03} leads to gauge parameters of the form
\begin{multline}
  \label{eq:eb}
  \veb^{(0)}=c_1 \Lt_{1} +\frac{3}{4}
  \left(
    \veb_{\Wt_{-2}}^{\prime} -\widetilde{\beta}  \veb_{\Wt_{-1}}  +\tfrac{8}{3} t c_1
  \right)\Lt_{0}
+  \left(
    \veb_{\Lt_{-1}} +\tfrac{3}{4} t 
    \left(
      \veb_{\Wt_{-2}}^{\prime} -\bt  \veb_{\Wt_{-1}} 
    \right)+t^{2}c_1
  \right)  \Lt_{-1}\\
  +c_2 \Wt_{2}
  + \left(
    \veb_{\Wt_{1}} +4t c_2
  \right) \Wt_{1}
  +
  \left(
    \veb_{\Wt_{0}} +3 t \veb_{\Wt_{1}} + 6 t^{2}c_2
  \right) \Wt_{0} \\
  + 
  \left(
    \veb_{\Wt_{-1}}  +2 t \veb_{\Wt_{0}}  +3 t^{2} \veb_{\Wt_{1}}  +4 t^{3} c_2
  \right)\Wt_{-1}
  +
  \left(
    \veb_{\Wt_{-2}} +t \veb_{\Wt_{-1}} +t^{2}\veb_{\Wt_{0}} +t^{3}\veb_{\Wt_{1}} +t^{4}c_2
  \right) \Wt_{-2}
\end{multline}
where $c_{1,2}$ are constant and $\veb_{\Lt_{-1}}$, $\veb_{\Wt_{1}}$, $ \veb_{\Wt_{0}}$, $ \veb_{\Wt_{-1}}$ and $ \veb_{\Wt_{-2}}$ are arbitrary functions of $\varphi$ which have to fulfill the conditions
\begin{subequations}\label{eq:ebcon}
  \begin{align}
    \veb_{\Wt_{1}}^{\prime} &=4 \bt  c_2 \label{eq:ebcon2}\\
    \veb_{\Wt_{0}}^{\prime} &=3 \bt  \veb_{\Wt_{1}}  \label{eq:ebcon3}\\
    \veb_{\Wt_{-1}}^{\prime} &=\tfrac{8}{3}c_1+2 \bt  \veb_{\Wt_{0}} \\
    \veb_{\Wt_{-2}}^{\prime \prime} &=\bt  
                                              \left(
                                              \veb^{\prime}_{\Wt_{-1}} +\tfrac{8}{3}c_1
                                              \right)+\bt^{\prime}  \veb_{\Wt_{-1}} -8 c_2 \, .\label{eq:ebcon4}
  \end{align}
\end{subequations}
Such transformations lead to a change $\widetilde\beta \to \widetilde\beta +\delta\widetilde\beta $ with
\begin{equation}
\delta\widetilde\beta =\veb^{\prime}_{\Lt_{-1}}  +\frac{3}{4}\left(-\widetilde\beta  \veb^{\prime}_{\Wt_{-2}} +\widetilde\beta ^{2} \veb_{\Wt_{-1}} +2 \veb_{\Wt_{1}} \right).
\end{equation}
Using \eqref{eq:ebcon} is easy to show that the corresponding boundary charges,
\begin{equation}
\delta \bar{\mathcal{Q}} \sim \oint \text{d}\vph c_1 \delta\widetilde\beta
\end{equation}
vanish, as integrands reduce to total derivatives,
\begin{equation}
 c_1 \delta\widetilde\beta = \frac{\dif}{\dif\vph} \left[ c_1 \veb_{\Lt_{-1}} 
+\frac{3c_1}{16c_2} \left( \veb_{\Wt_{1}} \veb_{\Wt_{-2}} + \widetilde\beta \veb_{\Wt_{1}} \veb_{\Wt_{-1}} + \frac{c_1}{3c_2} \veb_{\Wt_{1}}^2 \right) \right].
\end{equation}
 Thus, as already stated earlier, the function $\widetilde\beta $ is indeed pure gauge. A similar but more lengthy calculation shows that $\tilde w_1$ in \eqref{eq:ds} is pure gauge.

We have now completed the formulation of our theory and its boundary conditions. In the next section we perform several consistency checks to show that our theory is viable.

%==================================================

\section{Consistency of the theory}
\label{sec:consistency}

In this section we demonstrate the consistency of the theory defined in the previous section. Let us start by stating that the variational principle is well defined if we use a boundary term of the form~\cite{Gary:2012ms}
\begin{equation}
  \label{eq:Boundary}
  B=\frac{k}{4 \pi} \int_{\partial \mathcal{M}} \, \tr \left( A_{t}A_{\varphi}% -\bar{A}_{t} \bar{A}_{\varphi}
  \right)\dif t \dif \varphi \,.
\end{equation}
With this boundary term the first variation of the full action \eqref{eq:action} vanishes on-shell for all field variations that preserve our boundary conditions \eqref{eq:nolabel}-\eqref{eq:nw04}, $\delta I|_{\textrm{\tiny EOM}}=0$.

In the remainder of this section we perform more stringent consistency checks. In section \ref{sec:cbc} we construct the canonical boundary charges and show that they are finite, non-trivial, integrable and conserved. In section \ref{sec:ASA} we derive the asymptotic symmetry algebra, its central charges, and the consistency of the Jacobi identities in the presence of normal ordering. In section \ref{sec:unique} we show the uniqueness of the spin-connection, which then implies that the second order formulation is well-defined. In section \ref{sec:gauge-transf-half} we construct a global gauge transformation to a chiral half of an asymptotically AdS configuration, which makes our earlier observation of the chirality of the dual field theory more precise: it is a chiral CFT with Polyakov--Bershadsky symmetry algebra.

\subsection{Canonical boundary charges}\label{sec:cbc}

From canonical analysis~\cite{Banados:1994tn} we know that the canonical currents of the boundary charges are determined by
\begin{equation}\label{eq:nw05}
  \delta \mathcal{Q}\left[\ve\right] = \frac{k}{2\pi} \oint \dif\vph
  \tr\big(\ve \delta A_\vph\big) = \frac{k}{2\pi} \oint \dif\vph
  \tr\big(\ve^{(0)} \delta a_\vph^{(0)}\big)
\end{equation}
and similar for $\mathcal{\bar Q}$, where in the second equality cyclicity of the trace was used. 
Since only the field-independent gauge parameters $\ve_{\Lt_1}$,
$\ve_{\Wt_2}$, $\ve_{\Wt_1}$ and $\ve_{\Wt_0}$ contribute to this trace,
\begin{equation}
  \tr\big(\ve^{(0)} \delta a_\vph^{(0)}\big) = -4\ve_{\Lt_1}\delta
  l_{-1} + 9\ve_{\Wt_2} \delta w_{-2} - \frac{9}{4} \ve_{\Wt_1}\delta
  w_{-1} + \frac{3}{2} \ve_{\Wt_0}\delta w_{0}
\end{equation}
the canonical currents \eqref{eq:nw05} can easily be integrated 
%to
% \begin{equation}
%   \mathcal{Q}\left[\ve\right] = \frac{k}{2\pi} \oint \text{d}\vph
%   \left( -4 l_{-1} \ve_{\Lt_1} + 9 w_{-2} \ve_{\Wt_2} - \frac{9}{4} w_{-1}
%     \ve_{\Wt_1} + \frac{3}{2} w_{0} \ve_{\Wt_0} \right).
% \end{equation}
and rescaled
\begin{equation}
  \L_1 = -\frac{2k}{\pi}l_{-1}, \qquad \mathcal{W}_2 =
  \frac{9k}{2\pi}w_{-2}, \qquad
  \mathcal{W}_1 = -\frac{9k}{8\pi}w_{-1}, \qquad \mathcal{W}_0 =
  \frac{3k}{4\pi}w_{0}
\end{equation}
to yield the boundary charges
\begin{equation}\label{eq:nw08}
  \mathcal{Q}\left[\ve\right] = \oint \text{d}\vph \left( \L_{1}
    \ve_{L_1} + \mathcal{W}_{2} \ve_{\Wt_2} + \mathcal{W}_{1} \ve_{\Wt_1}
    + \mathcal{W}_{0} \ve_{\Wt_0} \right)\,.
\end{equation}
The asymptotic charges $\mathcal{Q}$ are finite, non-trivial, integrable in field space and conserved in time, $\partial_t \mathcal{Q}=0$. This is a fairly non-trivial indication that we have chosen meaningful boundary conditions when formulating our theory.

The asymptotic charges associated with the barred sector are trivial,
\begin{equation}
  \bar{\mathcal{Q}}\left[\bar{\ve}\right] = 0.
\end{equation}
Hence the theory we deal with is chiral, in the sense that it only has
one tower of asymptotic charges, $\mathcal{Q}$, while the barred tower
is trivial, $\bar{\mathcal{Q}}=0$.

\subsection{Asymptotic symmetry algebra}
\label{sec:ASA}

To establish the asymptotic symmetry algebra it is most convenient to
make use of the fact that for any state-dependent function $F$ on the phase space variations are canonically generated by the charges,
\begin{equation}
  - \delta_\ve F = \left\{ \mathcal{Q}\left[\ve\right] , F \right\}.
\end{equation}
This holds in particular also for variations of the charges themselves,
\begin{equation}\label{eq:nw07}
  - \delta_{\ve_1} \mathcal{Q}\left[\ve_2\right] = \left\{
    \mathcal{Q}\left[\ve_1\right] , \mathcal{Q}\left[\ve_2\right]
  \right\}.
\end{equation}
Evaluating the variations of the asymptotic charges on the left hand
side of \eqref{eq:nw07} gives the asymptotic symmetry algebra. To put
the algebra into a recognizable form, it is beneficial to combine
the asymptotic charges as
\begin{subequations}\label{eq:nw06}
\begin{align}
  \L & = \frac{2}{3}\mathcal{W}_2-\L_1 -\frac{2 \pi}{3 k}\mathcal{W}_0^2 = -\frac{2k}{\pi}\,\alpha^2 \\
  \J & = \frac{k}{2 \pi }-\frac{2}{3}\mathcal{W}_0\\
  \G_{\pm} & = \frac{2}{3} \left(\frac{3}{4}\L_1+\frac{3}{2}\mathcal{W}_0
             \mp \mathcal{W}_1 -\frac{3 k}{ 4\pi } \right).
\end{align}
\end{subequations}
After redefining the gauge parameters accordingly,
\begin{subequations}
\begin{align}
  \ve_\L & = \frac{3}{2}\ve_{\Wt_2}\\
  \ve_\J & = 3 \ve_{L_1}-\frac{3 \ve_{\Wt_0}}{2} + \left(
           \frac{9\pi}{2k}\J+ \frac{9}{4} \right)\ve_{\Wt_2}\\
  \ve_{\G_\pm} & = \ve _{\Lt_1} \mp \frac{3}{4}
                 \ve_{\Wt_1}+ \frac{3}{2} \ve_{\Wt_2},
\end{align}
\end{subequations}
the charge \eqref{eq:nw08} can be rewritten in terms of the new functions
\eqref{eq:nw06} as
\begin{equation}
  \mathcal{Q}\left[\ve\right] = \oint \text{d} \vph \left( \L \ve_\L +
    \J \ve_\J + \G_+ \ve_{\G_+} + \G_- \ve_{\G_-} \right).
\end{equation}
With these definitions, we find the variations
\begin{subequations}
\begin{align}
  \delta _{\L} \L & = - \L' \ve _{\L} -2 \L \ve _{\L}'-\frac{k}{4 \pi
                    } \ve _{\L}''' \\
  \delta _{\J}\L & =  -\J \ve _{\J}' \\ \displaybreak[1]
  \delta _{\G_\pm}\L & = -\frac{1}{2} \G_\pm' \ve _{\G_\pm}
                       -\frac{3}{2} \G_\pm \ve _{\G_\pm}' \\
  %
%  {}\nonumber\\
\displaybreak[1]
  \delta _{\L} \J & = - \J' \ve_\L - \J\ve_\L'  \\
\displaybreak[1]
  \delta _{\J} \J & = \frac{k}{3 \pi }\ve _{\J}' \\ 
\displaybreak[1]
  \delta _{\G_+} \J & = \pm \G_\pm \ve _{\G_\pm} \\
\displaybreak[1]
  % 
%  {}\nonumber\\
  % 
  \delta _{\L}\G_\pm & = -\G_\pm' \ve _{\L} -\frac{3}{2} \G_\pm \ve
                       _{\L}' \\
\displaybreak[1]
  \delta _{\J}\G_\pm & = \mp \G_\pm \ve _{\J} \\
\displaybreak[1]
  \delta _{\G_\pm}\G_\pm & = 0 \\
\displaybreak[1]
  \delta _{\G_\mp}\G_\pm & = \left(-\frac{3 }{2}\J' \mp \frac{6 \pi
                           }{k}\J^2 \mp \L\right)\ve _{\G_\mp} - 3 \J
                           \ve _{\G_\mp}' \mp \frac{k}{2 \pi }\ve
                           _{\G_\mp}''\,.
\end{align}
\end{subequations}
Inserting into \eqref{eq:nw07}, the Poisson bracket algebra is found
to be
\begin{subequations}
\begin{align}
  \left\{\L(\vph),\L(\bar{\vph}) \right\} & = \L'\delta-2\L\delta' -
                                            \frac{k}{4\pi}\delta'''\\
  \left\{\J(\vph),\L(\bar{\vph}) \right\} & = - \J \delta'\\
  \left\{\G_\pm(\vph),\L(\bar{\vph}) \right\} & = \frac{1}{2}
                                                \G_\pm'\delta -
                                                \frac{3}{2}\G_\pm
                                                \delta'\\
  \left\{\J(\vph),\J(\bar{\vph}) \right\} & = \frac{k}{3\pi} \delta'\\
  \left\{\J(\vph),\G_\pm(\bar{\vph}) \right\} & = \pm \G_\pm \delta\\
  \left\{\G_\pm(\vph),\G_\pm(\bar{\vph}) \right\} & = 0\\
  \left\{\G_\pm(\vph),\G_\mp(\bar{\vph}) \right\} & =
                                                    \left(\frac{3}{2}\J'\mp
                                                    \frac{6\pi}{k}\J^2
                                                    \mp \L\right)\delta - 3
                                                    \J\delta' \mp
                                                    \frac{k}{2\pi}\delta''
\end{align}
\end{subequations}
where $\delta' \equiv \partial_\vph \delta(\vph-\bar{\vph})$ and all
functions other than $\delta$ are functions of $\bar{\vph}$.

Writing the state-dependent functions in terms of their Fourier-modes,
\begin{subequations}
\begin{align}
  \L & = - \frac{1}{2\pi} \sum_{n\in \mathbb{Z}} \big(L_n -
       \frac{k}{4}\delta_{n,0}\big) e^{-i n\vph} \label{eq:needed} \\
  \J & = \frac{i}{2\pi} \sum_{n\in \mathbb{Z}} J_n e^{-i n\vph} \\
  \G_\pm & = - \frac{i^{\frac{1\mp1}{2}}}{2\pi} \sum_{n\in
           \mathbb{Z}+\frac{1}{2}} G_n^\pm e^{-i n\vph}
\end{align}
\end{subequations}
doing the same for $\delta(\vph - \bar{\vph})$ 
\begin{equation}
  \delta(\vph - \bar{\vph}) = \frac{1}{2\pi} \sum_{n\in \mathbb{Z}}
  e^{-i n(\vph-\bar{\vph})} = \frac{1}{2\pi} \sum_{n\in
    \mathbb{Z}+\frac{1}{2}} e^{-i n(\vph-\bar{\vph})}
\end{equation}
and replacing Dirac brackets by commutators as
$i\left\{\cdot,\cdot\right\} \rightarrow [\cdot,\cdot]$, one obtains
the semi-classical asymptotic symmetry algebra
\begin{subequations}
\begin{align}
  [L_n,L_m] & = (n-m)L_{n+m} + \frac{c}{12}n(n^2-1)\delta_{n+m,\,0}\\
  [L_{n},J_{m}] & =- m J_{n+m}\\
  [L_n,G^\pm_m] & = \big(\frac{n}{2}-m\big) G^\pm_{n+m}\\
  [J_n,J_m] & = -\frac{2k}{3} n\, \delta_{n+m,0}\\
  [J_n,G^\pm_m] & = \pm G^\pm_{n+m}\\
  [G^+_n,G^-_m] & = L_{n+m} + \frac{3}{2} (m-n) J_{n+m} +
                  \frac{3}{k}\sum_{p\in\mathbb{Z}} J_{m+n-p} J_p +
                  k\big(n^2-\frac{1}{4}\big) \delta_{n+m,\,0}
\end{align}
\end{subequations}
with $c=6k$. 

The Virasoro zero mode $L_0$ is related to the parameter $\alpha$ in the line-element \eqref{eq:nullusef} through
\eq{
L_0 -\frac k4 = 4k\alpha^2\,.
}{eq:stillnolabel}
In particular, $L_0$ is positive for positive level $k$ and $\alpha^2\geq 0$. This means that regular geometries (null warped AdS solitons) have positive energy $L_0>0$.

Taking into account normal ordering
\begin{equation}
  \sum_{p\in\mathbb{Z}} : J_{n-p}J_p : = \sum_{p\geq0} J_{n-p} J_p +
  \sum_{p<0} J_p J_{n-p}
\end{equation}
would make the semi-classical algebra inconsistent as the Jacobi identities would fail to hold. Demanding that they hold deforms some of the structure functions, exactly as in spin-3 Lobachevsky holography \cite{Afshar:2012nk}. Rescaling $G^\pm_n$ by a factor of $\sqrt{k-\frac{3}{2}}$ and defining $\hat{k} = - k - \frac{3}{2}$ the quantum asymptotic symmetry algebra
then takes the form of the quantum
Polyakov--Bershadsky algebra $\mathcal{W}_3^{(2)}$~\cite{Polyakov:1989dm, Bershadsky:1990bg},
\begin{subequations}
\label{eq:algebra}
\begin{align}
  [L_n,L_m] & = (n-m)L_{n+m} +
              \frac{\hat{c}}{12}n(n^2-1)\delta_{n+m,0}\\
  [L_{n},J_{m}] & =- m J_{n+m}\\
  [L_n,G^\pm_m] & = \big(\frac{n}{2}-m\big) G^\pm_{n+m}\\
  [J_n,J_m] & = \frac{2\hat{k}+3}{3} n \delta_{n+m,0}\\
  [J_n,G^\pm_m] & = \pm G^\pm_{n+m}\\
  [G^+_n,G^-_m] & = -(\hat{k}+3)L_{n+m} + \frac{3}{2}(\hat{k}+1) (n-m)
                  J_{n+m} \nonumber\\
            & \phantom{= } + 3 \sum_{p\in\mathbb{Z}} : J_{m+n-p} J_p :
              +3\frac{(\hat{k}+1)(2\hat{k}+3)}{2}\big(n^2-\frac{1}{4}\big)
              \delta_{n+m,0}
\end{align}
\end{subequations}
with central charge
\eq{
\hat{c} = -\frac{(2\hat{k}+3)(3\hat{k}+1)}{\hat{k}+3} = 6k + 16 + {\cal O}(1/k)\,.
}{eq:lalapetz}

As mentioned in section \ref{sec:null-warped-bcs}, the canonical charge in the barred sector, $\bar{\mathcal{Q}}$, is trivial. Thus, no asymptotic symmetry analysis needs to be performed in the barred sector. The asymptotic symmetry algebra associated with null warped AdS boundary conditions \eqref{eq:nolabel}-\eqref{eq:nw04} is therefore given by a single copy of $\mathcal{W}_{3}^{(2)}$ with central charge \eqref{eq:lalapetz}. This is one of our main results.

%==================================================

\subsection{Uniqueness of the spin connection}
\label{sec:unique}
As was pointed out in \cite{Lei:2015ika}, when considering gravitational theories in the first order formalism it can sometimes happen that the spin connection is not uniquely determined by the zuvielbein. In such cases the second order formulation is not well defined and it is therefore difficult to interpret the first order theory as a gravitational theory in the traditional sense. While this is not an obstruction to studying such theories, it can make the interpretation more difficult, and in particular labels such as ``null warped holography'' can become misleading. We check now that our theory does not have this issue.

The condition for having a well defined spin connection is that the generalized torsion condition
\begin{equation}\label{uniqueOmega}
  \dif e + e\wedge\omega + \omega\wedge e = 0
\end{equation}
has a unique solution for the connection $\omega$ in terms of the zuvielbein $e$~\cite{Fujisawa:2012dk}.

In the case under consideration, it is straightforward to check that the solution to \eqref{uniqueOmega} exists and is uniquely given by (the pure gauge functions are set to zero, i.e., $\widetilde w_{1}=\widetilde \beta=0$)
\begin{align}
\label{eq:omega}
  \omega =& \Big(e^\rho \Lt_1 +  e^{-\rho}l_{-1} \Lt_{-1} -\tfrac{2}{3} e^{2\rho}\Wt_2 + w_0 \Wt_0 + e^{-\rho}w_{-1} \Wt_{-1} + \left(\tfrac{2}{3}e^{2\rho} + e^{-2\rho}w_{-2}\right) \Wt_{-2}\Big)\, \dif\varphi\nonumber \\
  & + \tfrac{1}{2}e^\rho \Lt_{-1}\, \dif t\ .
\end{align}
Thus, our theory has a standard second-order interpretation in terms of metric and spin-3 field.

%==================================================

\subsection{Gauge transformation to half of AdS}
\label{sec:gauge-transf-half}
Given the fact that our asymptotic symmetry algebra is $\mathcal{W}_{3}^{(2)}$, it is natural to search for a gauge transformation to highest weight $\mathcal{W}_{3}^{(2)}$ boundary conditions~\cite{Campoleoni:2011hg,Ammon:2011nk}. These boundary conditions are most conveniently given in the diagonal embedding (for our conventions see Appendix \ref{sec:diagonal-embedding}).

We start by setting the pure gauge function, $\widetilde{w}_{1}$, in our connection~\eqref{eq:nw04} to zero and by inserting our redefinitions~\eqref{eq:nw06}. Furthermore we make a change of basis of the Lie algebra generators to the diagonal embedding after which we arrive at% , as this is the natural choice for a
% $\mathcal{W}_3^{(2)}$ symmetry,
\begin{align}
  \label{eq:adiag}
  a_\varphi & = -2\, \hat{\Lt}_1 +\frac{1}{27 k^2}
              \left( 18 k \pi 
              \left(
              -\L+\J
              \right) -27 \pi^{2} \J^{2}-6 \sqrt{6} k \pi 
              \left(
              \G_{+}+\G_{-}
              \right) + k^{2}
              \right)\hat{\Lt}_{-1}\nonumber\\
            & + \left(1-\frac{3 \pi  J }{k}\right)\Jt_0 -\sqrt{2}
              \left(
              \Gt^{-}_{+1/2}+ \Gt^{+}_{+1/2}
              \right) \nonumber\\
            & + \frac{1}{9 k} \left( 6 \sqrt{3} \pi  \G_{+} -\sqrt{2} (k-9
              \pi  \J)\right) \Gt^{-}_{-1/2} +\frac{1}{9k}
              \left(-6 \sqrt{3} \pi
              \G_{-}+\sqrt{2} (k-9 \pi  \J ) \right) \Gt^{+}_{-1/2}.
\end{align}
We are now searching for a gauge transformation $\lambda$ from our connection \eqref{eq:adiag} to a connection $a_{\varphi}^{\mathrm{HWG}}$ in highest weight gauge
\begin{equation}
  a_\varphi^{\mathrm{HWG}} = a_\varphi + \partial_\varphi \lambda +
  [a_\varphi,\lambda].
\end{equation}
Such a gauge transformation exists and is given by
\begin{equation}
\label{eq:gaugtr}
\lambda = -\frac{3}{2} \hat \Lt_{0}-\frac{1}{\sqrt{32}}
\left(
  \Gt^{-}_{-1/2}+\Gt^{+}_{-1/2}
\right).
\end{equation}
Since the gauge parameter is independent of state-dependent functions the asymptotic symmetry algebra should stay intact. The corresponding gauge transformed connection takes the form
\begin{align}
  a_\varphi^{\mathrm{HWG}} & = \hat{\Lt}_1% \nonumber\\
                        % &
                          +\underbrace{\frac{3\pi}{108 k}\left(-60 \L-42 \J -\tfrac{90 \pi}{k}\J^{2}-17 \sqrt{6} 
                          \left(
                          \G_{+}+\G_{-}
                          \right) + \tfrac{4 k }{3 \pi} \right)}_{{} =
                          -\frac{8\pi}{k}\left(\hat{\L}  -
                          \frac{6\pi}{k}\mathcal{U}^2\right)}
                          \hat{\Lt}_{-1} \nonumber\\
                        &+\underbrace{\left(\frac{1}{4}-\frac{3 \pi  \J
                          }{k}
                          \right)}_{{}=-\frac{12\pi}{k}\mathcal{U}}
                          \Jt_0 % \nonumber\\
                        % &
                          +\underbrace{\frac{1}{72 k} \left(
                          \sqrt{2}
                          \left(
                          5 k - 99 \pi \J
                          \right) -84\sqrt{3}  \G_{-}\right)
                          }_{{}=-\frac{8\pi}{k}\psi^{+}_{+}}
                          \Gt^{+}_{-1/2} \nonumber\\
                        &+\underbrace{\frac{1}{72 k} \left( 
                          -\sqrt{2}
                          \left(
                          5 k - 99 \pi \J
                          \right) +84 \sqrt{3}\G_{+} \right)
                          }_{{}=-\frac{8\pi}{k}\psi^{+}_{-}}
                          \Gt^{-}_{-1/2},
\end{align}
after redefining functions according to what is written beneath the braces. This connection has now all the dynamical components along highest weight generators and agrees with, e.g.,~ \cite[eq. (4.6)]{Bunster:2014mua} (for vanishing chemical potentials, where $A_{t}=0$).

This shows that our boundary conditions are equivalent to a chiral half of asymptotically AdS boundary conditions.
%================================================

\section{Thermodynamics}
\label{sec:thermo}

Having convinced ourselves that our theory is viable it is meaningful to discuss some of its physical properties. In this section we focus on thermodynamics. In order to be able to do this, we start by adding chemical potentials in section \ref{sec:adding-chem-potent}. Using these results, we then give a result for entropy of null warped AdS solitons in section \ref{sec:entropy}. In section \ref{sec:free} we compute the free energy and check the first law. In section \ref{sec:response} we calculate the response functions (or 1-point functions) and show that they coincide with the canonical charges, as expected for a consistent theory.

\subsection{Chemical potentials}
\label{sec:adding-chem-potent}

We  generalize our result by adding chemical potentials to the
theory. A standard way\footnote{For another possibility to introduce chemical potentials see~\cite{Gutperle:2011kf} and for a discussion of the differences between the approaches see~\cite{deBoer:2014fra}.}~\cite{Compere:2013gja,Compere:2013nba,Henneaux:2013dra,Bunster:2014mua} to do so is by looking for the most general
form of the time-component $A_t = b^{-1} \dif b + b^{-1} (a_t +
a_t^{(1)} ) b$ of the connection\footnote{For simplicity we will here
  and in the following use the notation $a_\mu = \hat{a}_\mu^{(0)}+
  a_\mu^{(0)}$.}, such that, without changing the form of $A_\vph$, the EOM are still satisfied. Thus we use a general ansatz
for $a_t$ and keep
\begin{equation}
\label{eq:aphichem}
  a_\vph  =  \Lt_1 - \tfrac{2}{3} \Wt_2 + l_{-1}(\vph,t)\Lt_{-1} + w_0(\vph,t) \Wt_0  + w_{-1}(\vph,t) \Wt_{-1}+ w_{-2}(\vph,t) \Wt_{-2}.
\end{equation}
However the state dependent functions can now in general also have a
time dependence. The EOM \eqref{eq:EOM} determine
 \begin{equation}
   a_t = a_t^{(\mu_\L)} + a_t^{(\mu_\J)} + a_t^{(\mu_{\G_+})} + a_t^{(\mu_{\G_-})}
 \end{equation}
with
\begin{subequations}
\begin{align}
  a_t^{(\mu_\L)} & = -\mu_\L \Lt_1 +\frac{1}{2} \mu_\L' \Lt_0 +
                   \frac{1}{2} \left( \frac{\pi}{k} 
                   (\G_+ + \G_-) + \frac{3\pi}{k}\J - \frac{1}{2}
                   \right) \mu_\L \Lt_{-1} \nonumber \\
                 & +\frac{2}{3}\mu_{\L} \Wt_2 + \left(\frac{2 \pi}{k}\J-1
                   \right) \mu_\L \Wt_0 + \frac{1}{3}\left(
                   \mu_\L' + \frac{2\pi}{k} (\G_- -
                   \G_+) \mu_\L \right) \Wt_{-1}  \nonumber \\
  \displaybreak[1]
                 & -\frac{1}{3}\left[ \frac{1}{4} \mu_\L'' +
                   \frac{\pi}{k} \left( \L
                   +\frac{3\pi}{2k}\J^2 + \frac{3}{2} \J
                   + \G_+ + \G_- - 
                   \frac{k}{8\pi} \right) \mu_\L \right] \Wt_{-2}\\
  {}\nonumber\\ \displaybreak[1]
  a_t^{(\mu_\J)} & = -\frac{1}{2} \mu_\J \Lt_{-1} - \frac{2}{3} \mu_\J
                   \Wt_0 +\frac{1}{6} \mu_\J \Wt_{-2}\\
  {}\nonumber\\
  a_t^{(\mu_{\G_\pm})} & =  \frac{1}{2} \mu_{\G_\pm} \Lt_1 \mp \frac{1}{2}
                         \mu_{\G_\pm} \Lt_0 + \frac{1}{4}\left[
                         \mp \mu_{\G_\pm}' + 3 \left(\frac{\pi}{k}  \J
                         + \frac{1}{2} \right) \mu_{\G_\pm} \right] \Lt_{-1}
                         \nonumber\\
                 & \mp \frac{2}{3} \mu_{\G_\pm} \Wt_1
                   + \mu_{\G_\pm} \Wt_0 - \frac{1}{3}
                   \left[\mu_{\G_\pm}' \mp \left(\frac{3\pi}{k}\J - \frac{1}{2}\right)
                   \mu_{\G_\pm} \right] \Wt_{-1} \nonumber\\
                 & + \frac{1}{6} \left[\pm
                   \mu_{\G_\pm}' - \left( \frac{\pi}{k}\G_\pm  +
                   \frac{3\pi}{k}\J + \frac{1}{2} \right) \mu_{\G_\pm}
                   \right] \Wt_{-2}.
\end{align}
\end{subequations}
The chemical potentials $\mu_\L$, $\mu_\J$, $\mu_{\G_+}$ and
$\mu_{\G_-}$  are in general arbitrary functions of $t$ and $\vph$. The EOM furthermore give conditions for the time
derivatives of the state-dependent functions,
\begin{subequations}\label{eq:nw11}
\begin{align}
  \dot{\L}  & = - \L' \mu_\L -2 \L \mu_\L' - \frac{k}{4 \pi }
              \mu_\L''' -\J  \mu_\J' - \frac{1}{2}(\G_+'  \mu_{\G_+} +
              \G_-'  \mu_{\G_-}) - \frac{3}{2}(\G_+  \mu_{\G_+}' +
              \G_-  \mu_{\G_-}')\\
  \dot{\J}  & = -\J'  \mu_\L - \J  \mu_\L'  +  \frac{k}{3 \pi}\mu_\J'
              +\G_+  \mu_{\G_+} - \G_-  \mu_{\G_-}\\
  \dot{\G_\pm}  & =  - \G_\pm'  \mu_\L -\frac{3}{2} \G_\pm  \mu_\L'
                  \mp \G_\pm  \mu_\J + \left( -\frac{3}{2} \J' \mp
                  \frac{6 \pi}{k} \J ^2 \mp \L\right) \mu_{\G_\mp} - 3
                  \J \mu_{\G_\mp}'  \mp \frac{k}{2\pi }\mu_{\G_\mp}''.
\end{align}
\end{subequations}
These encode the asymptotic symmetry algebra. In the case of vanishing chemical potentials, equations \eqref{eq:nw11}
reduce to
\begin{equation}
  \dot{\L} = \dot{\J}= \dot{\G}_\pm = 0
\end{equation}
such that $a_\varphi$ no longer depends on $t$ and one recovers the
results from section \ref{sec:null-warped-bcs}.

We will in the following only consider zero mode solutions with
constant chemical potentials.
Furthermore, we will work in the Euclidean framework with $\tau = it$ where $\tau \sim \tau+1$,
such that $a_\tau = - i a_t$. The charges and chemical potentials
are not affected by this change. Note that our charges and chemical potentials are real in Euclidean signature, as there is no barred sector to form their imaginary part. This differs from the situation for black holes in the non-principal embedding with AdS boundary conditions (see for example \cite{Bunster:2014mua}).

%==================================================

\subsection{Entropy = chiral Cardy}\label{sec:entropy}
In this section we determine entropy along the lines of~\cite{Bunster:2014mua}. It is given by
\begin{equation}\label{eq:entropy}
S = k \tr( a_t a_{\vph} ) = 2 \pi \big( 2 \mu_\L \L +
  \mu_\J \J  + \frac{3}{2} (\mu_{\G_+} \G_+ + \mu_{\G_-} \G_+) \big)\,.
\end{equation}
Furthermore we find holonomy conditions
\begin{multline}
\label{eq:hololong}
\frac{k^2}{216 \pi ^3} \left(2 k \mu_\J^3-27 \pi  \mu_{\G_-}
  \mu_{\G_+} (\G_- \mu_{\G_-}+\G_+ \mu_{\G_+}+2 \J
  \mu_\J)\right)+\frac{\mu_\L^3 }{2 \pi } \left(-2 \pi  \J^3+\G_- \G_+
  k-\J k \L\right)\\
+\frac{k \mu_\L^2 }{12 \pi ^2} (3 \pi  \J (3 \G_- \mu_{\G_-}+3 \G_+
\mu_{\G_+}+2 \J \mu_\J)+2 k \mu_\J \L) \\
-\frac{k \mu_\L }{12 \pi ^2}
\left(-18 \pi  \J^2 \mu_{\G_-} \mu_{\G_+}+\J k \mu_\J^2+3 k \mu_{\G_-}
  \mu_{\G_+} \L\right) = 0
\end{multline}
and
\begin{equation}
\label{eq:holomedium}
\frac{2 \pi  \mu_\L }{k} (3 \G_- \mu_{\G_-}+3 \G_+ \mu_{\G_+}+2 \J
\mu_\J)+\frac{12 \pi }{k}  \J \mu_{\G_-} \mu_{\G_+}+\frac{4 \pi
  \mu_\L^2 \L}{k}-\frac{2 \mu_\J^2}{3} + 8\pi^2 = 0.
\end{equation}
Additional conditions are given by the EOM \eqref{eq:nw11} for
constant charges and chemical potentials, 
\begin{subequations}
\label{eq:holoEOM}
  \begin{align}
    \G_+ \mu_{\G_+}-\G_- \mu_{\G_-} &= 0\\
    \mu_{\G_+} \L +\frac{6 \pi  \J^2 \mu_{\G_+}}{k} + \G_- \mu_\J &= 0\\
    \mu_{\G_-} \L +\frac{6 \pi  \J^2 \mu_{\G_-}}{k} + \G_+ \mu_\J &= 0 \, .
  \end{align}
\end{subequations}
Up to redefinitions
\begin{subequations}
  \begin{align}
    \L = -4 \hat{\L}, \quad \J = -4 \mathcal{U}, \quad \G_{\pm} = \pm 4
    \psi_\pm\\
    \mu_\L = -\xi, \quad \mu_\J = \nu \quad \mu_{\G_\pm} = \mp \vartheta_\pm
  \end{align}
\end{subequations}
these equations are equivalent to the holonomy conditions and field equations
obtained in \cite{Bunster:2014mua}. We can thus write the chemical
potentials in terms of the boundary charges by using the results of
\cite{Bunster:2014mua}, such
that the entropy \eqref{eq:entropy} is given by
\begin{equation}
\label{eq:entropyfull}
S = 2\pi \sqrt{-8\pi k\L} \,\cos\left(
    \tfrac{\Phi}{3}  \right) 
\end{equation}
where
\begin{equation}
  \label{eq:PHI}
  \Phi=\arcsin\Big( 3 \sqrt{-\frac{6\pi^3}{k^3\L^{3}}}
      \big[ -\J^3 - \frac{k}{2\pi}\big( \J\L
        - \G_+\G_-\big) \big]\Big) \,.
\end{equation}
This is well defined if $\L \leq 0$ and 
\begin{equation}
\bigg|3 \sqrt{-\frac{6\pi^3}{k^3\L^{3}}}
      \Big( \J^3 + \frac{k}{2\pi}\big( \J\L
        - \G_+\G_-\big) \Big)\bigg| \leq 1\,.
\end{equation}
Note that the condition $\L \leq 0$ is equivalent to non-negative $\alpha^2$ in \eqref{eq:nullusef}, so the physical states whose entropy we are calculating are null warped AdS solitons in the classification of section \ref{sec:null-warped-ads}.

In the special case of vanishing lower-spin charges, $\J=\G_+=\G_-=0$, entropy simplifies to
\eq{
S = 2\pi \sqrt{\frac{ 4 c E}{6}} = 4\pi k |\alpha|
}{eq:Ssimple}
where we defined the energy $E=-2\pi \L=L_0-\tfrac k4$ in the middle equation [the factor and sign originate from \eqref{eq:needed}; see also \eqref{eq:stillnolabel}], which is of chiral Cardy-type. The reason for the four in front the central charge is due to the fact that our central charge in equation \eqref{eq:Ssimple} is a quarter of the standard central charge, $4 c = c_{\mathrm{S}}=\frac{3 \ell}{2 G}$~\cite{Ammon:2011nk}. With the standard central charge we arrive at the standard Cardy formula. This provides yet-another consistency check of our theory: the macroscopic entropy matches the microscopic one. 

\subsection{First law and free energy}\label{sec:free}

The equations for the chemical potentials
\begin{subequations}
    \label{eq:substi}
  \begin{align}
    \mu_{\L}&=2 \pi  \sqrt{\frac{k}{E}}  \left(2  \pi  \mathcal{J}
              \sqrt{\frac{3}{E k}} \frac{\sin\left(\frac{\Phi}{3}\right)}{\cos(\Phi)}
              -\frac{\cos\left(\frac{2 \Phi }{3}\right)}{\cos(\Phi)}\right)
    \\
    \mu_{\J}&=-2 \sqrt{3} \pi  \frac{\sin\left(\frac{\Phi }{3}\right)}{\cos(\Phi)} 
              \left(1-\frac{12 \pi ^2 \mathcal{J}^2}{E k}\right)\\
    \mu_{\G_{+}}&=-\frac{4 \sqrt{3} \pi ^2 \G_{-}\sin \left(\frac{\Phi
                  }{3}\right)}{E \cos(\Phi)}\\
    \mu_{\G_{-}}&=-\frac{4 \sqrt{3} \pi ^2 \G_{+}\sin \left(\frac{\Phi
                  }{3}\right)}{E \cos(\Phi)}
  \end{align}
\end{subequations}
together with \eqref{eq:PHI} fulfill the holonomy conditions and lead to \eqref{eq:entropyfull}.

We use $E=-2\pi\L$ to write the entropy as a function of $E$, i.e., $S=S(E,\J,\G_{\pm})$. Now the first law of thermodynamics 
\begin{align}
  \label{eq:Sdiff}
  \dif S&=\beta \dif E-\beta (\Omega_{\J}\dif \J+\Omega_{\G_{+}}\dif \G_{+}+\Omega_{\G_{-}}\dif \G_{-})
\end{align}
holds provided that
\begin{subequations}
  \begin{align}
    \label{eq:provS}
    \beta&:=\left( \frac{\partial S}{\partial E}\right)_{\J, \G_\pm}=-\mu_{\L}\\
    \beta \Omega_{\J}&:=-
                       \left(
                       \frac{\partial S}{\partial \J}
                       \right)_{E,\G_{\pm}}=-2\pi \mu_{\J}\\
    \beta \Omega_{\G_{\pm}}&:=-
                             \left(
                             \frac{\partial S}{\partial \G_{\pm}}
                             \right)_{E,\J,\G_{\mp}}=- 2 \pi \mu_{\G_{\pm}} \,.
  \end{align}
\end{subequations}
For $\J=\G_+=\G_-=0$ we get
\begin{equation}
  \label{eq:betazero}
  T=\beta^{-1}=\frac{\sqrt{E}}{2 \pi \sqrt{k}} \, .
\end{equation}
We can now determine the free energy $F$ 
\begin{equation}
  \label{eq:Fgen}
    F(T,\Omega_{\J},\Omega_{\G_{\pm}})=E-T S - \Omega_{\J}\J -\Omega_{\G_{+}}\G_{+} -\Omega_{\G_{-}}\G_{-}=-E+\frac{1}{2}  \left(  \Omega_{\G_{+}}\G_{+}+\Omega_{\G_{-}}\G_{-}\right) \, .
\end{equation}

We can solve the holonomy conditions \eqref{eq:holomedium} and \eqref{eq:holoEOM} for $E,\J,\G_{+},\G_{-}$. For
\begin{equation}
  \label{eq:EnotInf}
  \Omega_{\J}= \frac{3}{2 \pi} \Omega_{\G_{+}}\Omega_{\G_{-}}
\end{equation}
there are two real free energy branches. Alternatively if the inequality 
\begin{equation}
  \label{eq:EnotInf}
  \Omega_{\J}\neq \frac{3}{2 \pi} \Omega_{\G_{+}}\Omega_{\G_{-}}.
\end{equation}
holds we can solve three of the holonomy conditions to get a remaining quartic equation in $\J$. This means that for generical non-vanishing chemical potentials and positive temperature one gets four, possibly complex, solutions for the free energy. In the regime where the temperature is smaller than
\begin{equation}
  \label{eq:Tmax}
  T_{\mathrm{max}}=\frac{\sqrt{-3+2\sqrt{3}}(\Omega_{\J}-\frac{3}{2 \pi} \Omega_{\G_{+}} \Omega_{\G_{-}})^{2}}{24 \pi | \Omega_{\G_{+}} \Omega_{\G_{-}}| }
\end{equation}
there are four distinct free energies $F_{1}$ to $F_{4}$.  When the temperature is equal to $T_{\mathrm{max}}$ two of them, $F_{1}$ and $F_{2}$, coincide. For temperatures higher than $T_{\mathrm{max}}$ these two branches get complex and are thus no viable branches anymore. 

\begin{figure}[h!]
  \centering
  \begin{tabular}[h]{l l}
    \includegraphics[height=4cm]{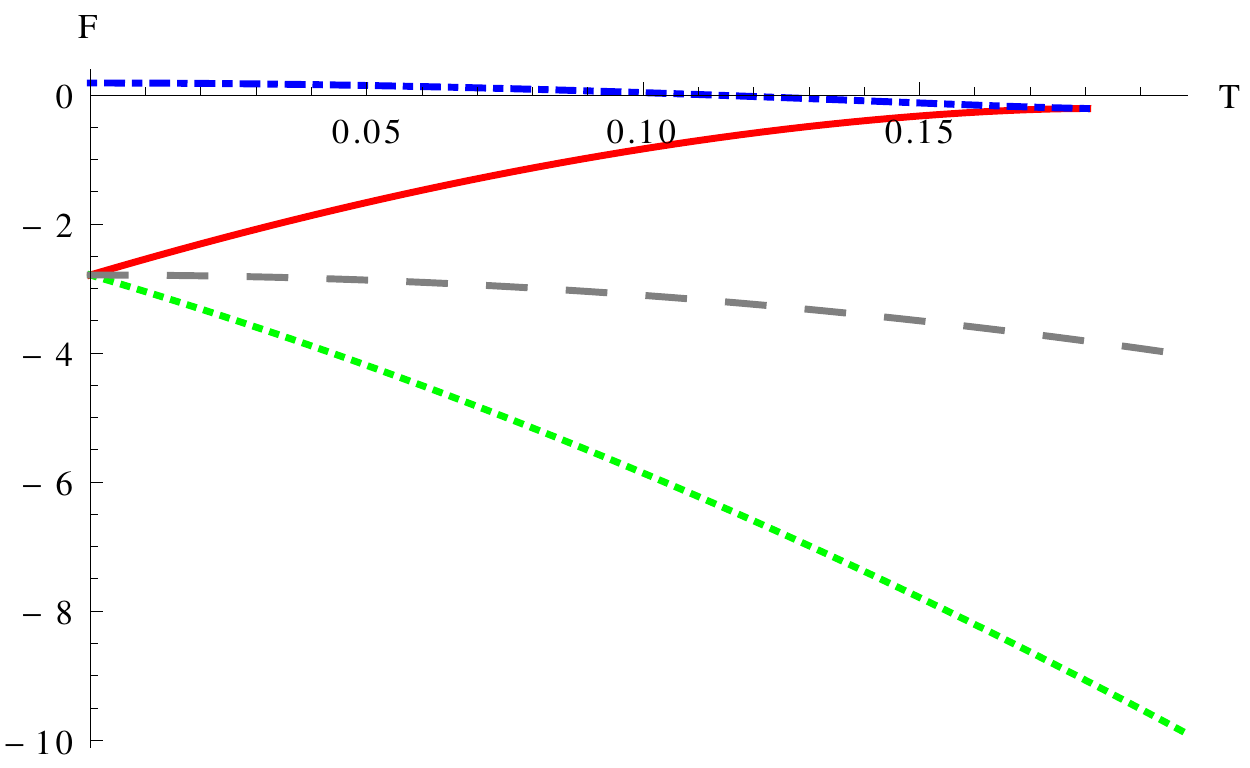} & \includegraphics[height=4cm]{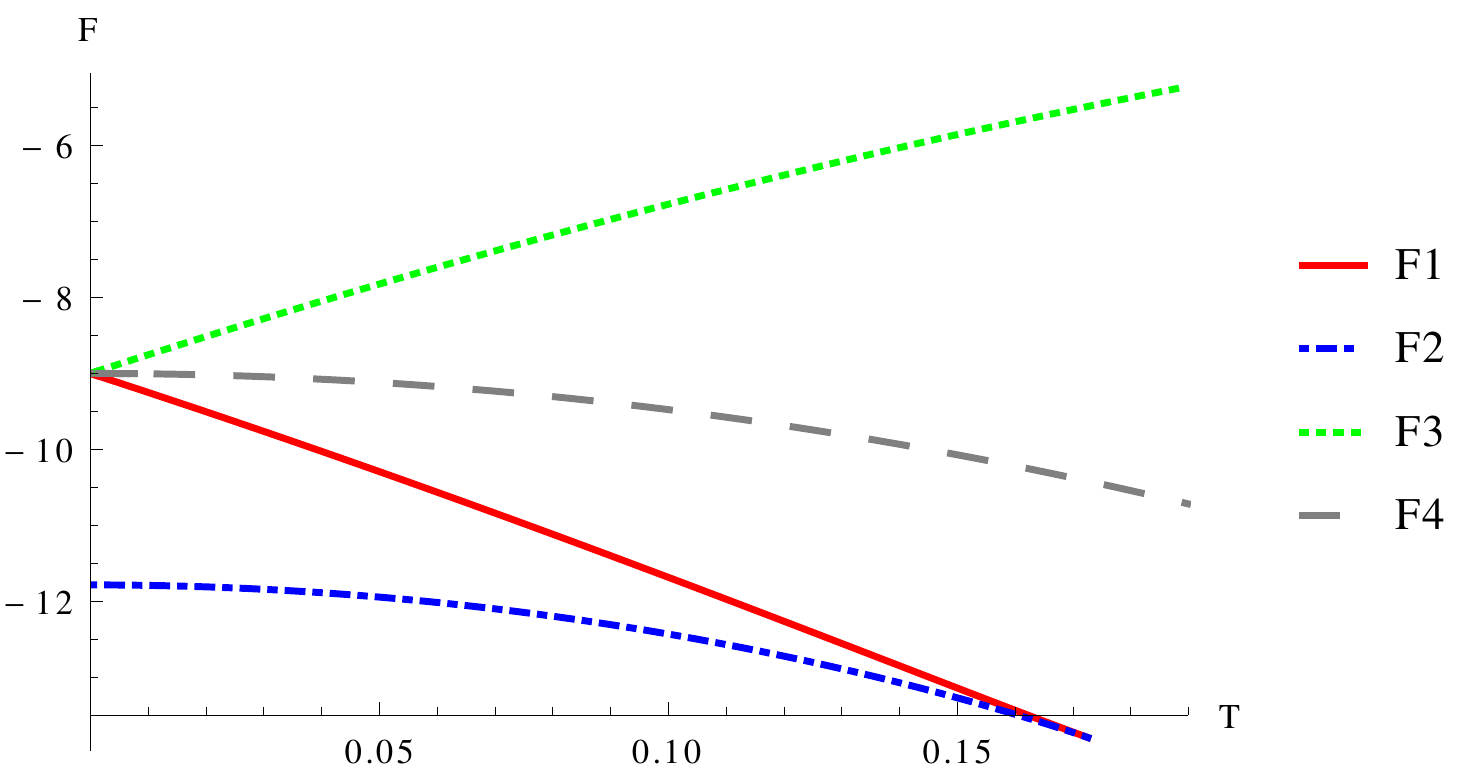}
  \end{tabular}
  \caption{For $\Omega_{\J}< \frac{3}{2 \pi} \Omega_{\G_{+}}\Omega_{\G_{-}}$ the $F_{3}$ branch has the lowest free energy, whereas for $\Omega_{\J}> \frac{3}{2 \pi} \Omega_{\G_{+}}\Omega_{\G_{-}}$ and temperatures lower than $T_{\mathrm{max}}$ the branch $F_{2}$ dominates.
We set $k=1$,  $\Omega_{\G_{+}}=8$ and  $\Omega_{\G_{-}}=4$. Left: $-10=\Omega_{\J}< \frac{3}{2 \pi} \Omega_{\G_{+}}\Omega_{\G_{-}}$. Right: $40=\Omega_{\J}> \frac{3}{2 \pi} \Omega_{\G_{+}}\Omega_{\G_{-}}$. }
  \label{fig:F}
\end{figure}

We will now restrict our discussion to  $\Omega_{\G_{\pm}} >0$. The $F_{1}$ branch is the only branch which is well defined for vanishing  $\Omega_{\G_{\pm}}$. Its small $T$ expansion is given by
\begin{equation}
  \label{eq:fonesmallT}
  F_{1}= -\frac{k}{48 \pi^{4}} 
                    \left(
                    3 \Omega_{\G_{+}}^{2}\Omega_{\G_{-}}^{2}-6 \pi \Omega_{\G_{+}}\Omega_{\G_{-}} \Omega_{\J}+ 4 \pi^{2}\Omega_{\J}^{2}
                    \right)+
                    \left(
                      \frac{3 \Omega_{\G_{+}} \Omega_{\G_{-}}}{2 \pi}-\Omega_{\J}
                    \right) k T + O\left(T^{2}\right) \, .
\end{equation}
At zero temperature $F_{1}$ coincides with the branches $F_{3}$ and $F_{4}$ whereas $F_{2}$ has generically a different zero temperature value
\begin{equation}
  \label{eq:ftwosmallT}
    F_{2}=-\frac{k \Omega_{\J}^{3}}{54 \pi \Omega_{\G_{+}}\Omega_{\G_{-}}}-\frac{4 \pi^{2} \Omega_{\J} k T^{2}}{\Omega_{\J}-\frac{3}{2 \pi} \Omega_{\G_{+}} \Omega_{\G_{-}}}+O\left(T^{3}\right) \, .
\end{equation}
For large temperatures the branches $F_{1}$ and $F_{2}$ vanish and only the other two remaining branches dominate with a large $T$ behavior given by
\begin{equation}
  \label{eq:FTinfinit}
    F_{3,4}=-4 k \pi^{2} T^{2} + O(T) \, .
\end{equation}
We summarize this discussion in the following table (for more details see  figure \ref{fig:F})
\begin{center}
  \begin{tabular}{|c|c|c|c|}
    \hline
   $T=0$ &$0 < T < T_{\mathrm{max}}$ &    $T = T_{\mathrm{max}}$ &  $T_{\mathrm{max}}<T$\\
    \hline
    $F_{1}=F_{3}=F_{4}, F_{2}$ &$F_{1},F_{2},F_{3},F_{4}$ & $F_{1}=F_{2},F_{3},F_{4}$ & $F_{3},F_{4}$\\
    \hline
  \end{tabular} .
\end{center}
Scanning the free energy for all values of $T$, $\Omega_\J$ and $\Omega_{\G_\pm}$ we have found no first or second order phase transitions between the four branches.

\subsection{Holographic response functions}\label{sec:response}

In the presence of chemical potentials the variation of the $t$-component of the connection no longer vanishes, $\delta A_t \neq 0$. As a consequence, the boundary term \eqref{eq:Boundary} no longer yields a well-defined variational principle and has to be replaced by
\begin{equation}
B = \frac{k}{4\pi}
  \int_{\partial \mathcal{M}} \tr \left((a_t+\widetilde{a}_t)
    a_\vph\right) \,\dif t \,\dif\vph
\end{equation}
with
\begin{multline}
\widetilde{a}_t = \frac{1}{2}\left( 4\mu_\L - \mu_{\G_+} - \mu_{\G_-}
\right) \Lt_1 - \left( \mu_{\G_+} + \mu_{\G_-} + \frac{3\pi}{4k}\J
  \mu_\J \right) \Lt_{-1} - \frac{4}{3}\mu_\L \Wt_2\\
 + \frac{2}{3} \left( \mu_{\G_+} - \mu_{\G_-} \right) \Wt_1
 + \left( 2\mu_\L - \mu_{\G_+} - \mu_{\G_-} - \frac{2\pi}{k}\J\mu_\L
 \right) \Wt_0 + \frac{1}{4} \mu_\L \Wt_{-2}\,.
\end{multline}
The full action \eqref{eq:action} with the boundary term above has a well-defined variational principle, i.e., if the chemical potentials are fixed the variation of the action vanishes on-shell. 

If the chemical potentials are allowed to vary according to the holographic dictionary this corresponds to switching on non-normalizable modes or, equivalently, sources. In this situation the first variation of the full action yields the response functions
\begin{equation}
\big(\delta I_{\textrm{CS}} + \delta B\big)\big|_{\textrm{\tiny EOM}} = \int_{\partial \mathcal{M}} \big( \L\delta\mu_\L +
\J\delta\mu_\J + \G_+\delta \mu_{\G_+} + \G_-\delta \mu_{\G_-} \big) \,\dif t \,\dif \vph\ .
\label{eq:response}
\end{equation}
The response functions coincide precisely with the canonical boundary charges, which provides the final consistency check of our theory in the present work.

%==================================================

\section{Discussion}
\label{sec:discussion}

We have provided a simple example of a higher spin theory that has a chiral asymptotic symmetry algebra, namely a single copy of the Polyakov--Bershadsky algebra. Geometrically we were motivated by null warped AdS, but as we showed there is an equivalent interpretation as a chiral half of AdS. We conclude now with relations to null warped AdS in topologically massive gravity and further comments on our results.

In topological massive gravity a natural choice of boundary conditions leads to an asymptotic symmetry algebra that consists of a single copy of the Virasoro algebra and a $\widehat{u(1)}$-current algebra~\cite{Anninos:2010pm}. % [eq. (4.2)]
Even though the boundary conditions are consistent they allow the existence of normalizable polynomial modes that grow linearly in time. This is problematic since it produces closed timelike curves at either early or late times and therefore more restrictive boundary conditions where proposed~\cite{Anninos:2010pm}. % [eq. (4.3)]
The asymptotic symmetry algebra of the theory subject to these more restrictive boundary conditions is chiral, like our higher spin theory. The authors remark that their analysis indicates that the classical theory of topologically massive gravity in null warped AdS %with both boundary conditions 
is chiral. This is in agreement with our analysis, which also leads us to a chiral theory. Also our result for entropy \eqref{eq:Ssimple} is similar to the result for null warped black hole entropy in topologically massive gravity \cite{Anninos:2010pm}: it scales linearly with the inverse gravitational coupling $k$ and the state-dependent parameter $\alpha$.

While unitarity is not necessary for holography \cite{Grumiller:2008qz,Grumiller:2013at,Vafa:2014iua}, it is still important to know under which conditions (if any) a given theory is compatible with unitarity. For our theory the same remarks apply as for Lobachevsky holography \cite{Afshar:2012nk}, see also the general discussion of unitarity bounds \cite{Castro:2012bc}: the algebra \eqref{eq:algebra} allows for non-trivial unitary highest-weight representations only for $\hat c=1$. For any other value of the central charge the theory has no unitary highest weight representations, unless we either add or truncate further degrees of freedom.

It would be interesting to holographically compute entanglement entropy \cite{Ryu:2006bv} for our theory, along the lines of \cite{Ammon:2013hba,deBoer:2013vca}. While originally these computations were mainly devised for AdS-holography, they appear to work also for flat space holography \cite{Bagchi:2014iea} and possibly more generally for non-AdS holography. This may shed additional light on the interpretation of our theory in terms of null warped AdS versus interpreting it as a chiral half of AdS.

Finally, let us mention another possible check of null warped AdS higher spin holography that should be feasible. In the present work we have calculated holographically the 1-point functions. As exploited recently for flat space holography \cite{Bagchi:2015wna}, knowing the 1-point functions on an arbitrary background allows to iteratively construct all the higher $n$-point functions, at least for theories that have a Chern--Simons formulation. Since this applies to our theory we believe it should be possible to extend our checks in the present work to arbitrary $n$-point functions. We intend to consider this in future work.

%==================================================

\section*{Acknowledgments}
\label{sec:acknowledgments}
We thank Eric Perlmutter for collaboration during the early stages of this project.
We are grateful to our previous collaborators on non-AdS holography in higher spin gravity for discussions.

This work was supported by the START project Y 435-N16 of the Austrian Science
Fund (FWF) and the FWF projects I 952-N16, I 1030-N27 and P 27182-N27. During the
final stages MG and SP were supported by the FWF project P 27396-N27 and DG by the program
Science without Borders, project CNPq-401180/2014-0.

%==================================================

\appendix

\section{Principal embedding}
\label{sec:principal-embedding}

We use a principal embedding of $sl(2,\mathbb{R})$ into
$sl(3,\mathbb{R})$ 
\begin{subequations}
  \begin{align}
    [\Lt_i,\Lt_j] & = (i-j)\Lt_{i+j}\\
    [\Lt_i,\Wt_m] & = (2i-m) \Wt_{i+m}\\
    [\Wt_m,\Wt_n] & = -\frac{3}{16}(m-n)(2 m^2+2 n^2-mn-8) \Lt_{m+n}
  \end{align}
\end{subequations}
where $i,j=-1,0,1$ and $m,n=-2,-1,0,1,2$.
An explicit realization is given by
\begin{align}
  \Lt_1  & = \left(\begin{array}{ccc}	0 & 0 & 0 \\
                   -\sqrt{2} & 0 & 0 \\ 0 & -\sqrt{2} &
                                                        0 \end{array}\right) &
  \Lt_0 & = \left(\begin{array}{ccc}	1 & 0 & 0 \\ 0 & 0 & 0 \\ 0 &
                                                                      0 & -1 \end{array}\right) &
  \Lt_{-1} & = \left(\begin{array}{ccc}	0 & \sqrt{2} & 0 \\ 0 & 0 &
                                                                    \sqrt{2}
                     \\ 0 & 0 & 0 \end{array}\right) \nonumber\\
  \Wt_2 & = \left(\begin{array}{ccc}	0 & 0 & 0 \\ 0 & 0 & 0 \\ 3 &
                                                                      0
                                              & 0 \end{array}\right) &
  \Wt_1 & = \left(\begin{array}{ccc}	0 & 0 & 0 \\
                -\frac{3}{2\sqrt{2}} & 0 & 0 \\ 0 &
                                                    \frac{3}{2\sqrt{2}} & 0 \end{array}\right) &
  \Wt_0 & = \left(\begin{array}{ccc}	\frac{1}{2} & 0 & 0 \\ 0 & -1
                                                        & 0 \\ 0 & 0 &
                                                                       \frac{1}{2}	\end{array}\right)\nonumber\\
  \Wt_{-1} & = \left(\begin{array}{ccc}	0 & \frac{3}{2\sqrt{2}} & 0 \\
                   0 & 0 & -\frac{3}{2\sqrt{2}} \\ 0 & 0 &
                                                           0 \end{array}\right) &
  \Wt_{-2} & = \left(\begin{array}{ccc}	0 & 0 & 3 \\ 0 & 0 & 0 \\ 0 &
                                                                      0 & 0 \end{array}\right).
\end{align}
and the only non-vanishing traces are
\begin{align}
  \tr(\Lt_0 \Lt_0) & = 2 & \tr(\Lt_1 \Lt_{-1}) & = -4 &  && \nonumber\\
  \tr(\Wt_0 \Wt_0) & = \frac{3}{2}& \tr(\Wt_1 \Wt_{-1}) & = -\frac{9}{4}&\tr(\Wt_2 \Wt_{-2}) & = 9 \, .
\end{align}

%==================================================

\section{Diagonal embedding}
\label{sec:diagonal-embedding}
The diagonal embedding of $sl(2,\R)$ into $sl(3,\R)$ is  given by\footnote{For easier comparison of the connections in Section \ref{sec:gauge-transf-half} we follow the conventions of \cite{Bunster:2014mua}.}
\begin{subequations}
  \begin{align}
    \label{eq:diagALG}
    [\hat{\Lt}_{i},\hat{\Lt}_{j}]&=(i-j) \hat{\Lt}_{i+j}\\
    [\hat{\Lt}_{i},\Jt_{0}]&=0\\
    [\hat{\Lt}_{i},\Gt_{m}^{a}]&=\left( \frac{i}{2}-m \right) \Gt_{i+m}^{a}\\
    [\Gt_{m}^{+},\Gt_{n}^{-}]&=\hat{\Lt}_{m+n}-\frac{3}{2}(m-n) \Jt_{0}\\
  [\hat{\Jt}_{0},\Gt_{m}^{a}]&=a \Gt_{m}^{a}
  \end{align}
\end{subequations}
where $i,j=-1,0,1$, $m,n=-1/2,1/2$ and $a=1,-1$.
An explicit realization is given by
\begin{align}
  \hat{\Lt}_1  & = \left(\begin{array}{ccc}	0 & 0 & 0 \\
                         0 & 0 & 0 \\ 1 & 0 &  0 \end{array}\right) &
  \hat{\Lt}_0 & = \left(\begin{array}{ccc}	\frac{1}{2} & 0 & 0 \\
                        0 & 0 & 0 \\ 0 &  0 &
                                              -\frac{1}{2} \end{array}\right) &
  \hat{\Lt}_{-1} & = \left(\begin{array}{ccc}	0 & 0 & -1 \\
                           0 & 0 & 0 \\ 0 & 0 & 0 \end{array}\right)
                                                \nonumber\\
  \Jt_0 & = \left(\begin{array}{ccc}	\frac{1}{3} & 0 & 0 \\
                  0 & -\frac{2}{3} & 0 \\ 0 & 0 &
                                                  \frac{1}{3} \end{array}\right) &
  \Gt^{+}_{+1/2} & = \left(\begin{array}{ccc}	0 & 0 & 0 \\
                           0 & 0 & 0 \\ 0 & 1 & 0 \end{array}\right) &
  \Gt^{+}_{-1/2} & = \left(\begin{array}{ccc}	0 & 1 & 0 \\
                           0 & 0 & 0 \\ 0 & 0 &
                                                0	\end{array}\right)\nonumber\\
  \Gt^{-}_{+1/2} & = \left(\begin{array}{ccc}	0 & 0 & 0 \\
                           1 & 0 & 0 \\ 0 & 0 & 0 \end{array}\right) &
  \Gt^{-}_{-1/2} & = \left(\begin{array}{ccc}	0 & 0 & 0 \\
                           0 & 0 & -1 \\ 0 & 0 & 0 \end{array}\right)
\end{align}
and the only non-vanishing traces are
\begin{align}
  \label{eq:Diagtr}
 \tr(\hat{\Lt}_{0}\hat{\Lt}_{0})&=\frac{1}{2} & \tr(\hat{\Lt}_{1}\hat{\Lt}_{-1})&=-1 & \nonumber \\
 \tr(\Jt_{0}\Jt_{0})&=\frac{2}{3} &  \tr(\Gt_{+1/2}^{+} \Gt^{-}_{-1/2})&=-1 &  \tr(\Gt_{-1/2}^{+} \Gt_{+1/2}^{-})&=1 \, .
\end{align}

%==================================================

\addcontentsline{toc}{section}{References}

\bibliography{nullwarped}

\providecommand{\href}[2]{#2}\begingroup\raggedright\begin{thebibliography}{10}

\bibitem{BBB}
V.~Balasubramanian, A.~Maloney, D.~Marolf, and J.~Simon, ``Bits, branes, black
  holes,'' May, 2012.
\newblock KITP program and workshop, Santa Barbara, see numerous talks at the
  \href{http://www.kitp.ucsb.edu/activities/dbdetails?acro=bitbranes12}{program
  and conference webpage}.

\bibitem{'tHooft:1993gx}
G.~'t~Hooft, ``{Dimensional reduction in quantum gravity},'' in {\em {Salamfest
  1993:0284-296}}, pp.~0284--296.
\newblock 1993.
\newblock
\href{http://www.arXiv.org/abs/gr-qc/9310026}{{\tt gr-qc/9310026}}.
\newblock
%%CITATION = GR-QC/9310026;%%.

\bibitem{Susskind:1994vu}
L.~Susskind, ``{The World as a hologram},'' {\em J.Math.Phys.} {\bf 36} (1995)
  6377--6396,
\href{http://www.arXiv.org/abs/hep-th/9409089}{{\tt hep-th/9409089}}.
%%CITATION = HEP-TH/9409089;%%.

\bibitem{Maldacena:1997re}
J.~M. Maldacena, ``{The Large N limit of superconformal field theories and
  supergravity},'' {\em Adv.Theor.Math.Phys.} {\bf 2} (1998) 231--252,
\href{http://www.arXiv.org/abs/hep-th/9711200}{{\tt hep-th/9711200}}.
%%CITATION = HEP-TH/9711200;%%.

\bibitem{Gubser:1998bc}
S.~Gubser, I.~R. Klebanov, and A.~M. Polyakov, ``{Gauge theory correlators from
  noncritical string theory},'' {\em Phys.Lett.} {\bf B428} (1998) 105--114,
\href{http://www.arXiv.org/abs/hep-th/9802109}{{\tt hep-th/9802109}}.
%%CITATION = HEP-TH/9802109;%%.

\bibitem{Witten:1998qj}
E.~Witten, ``{Anti-de Sitter space and holography},'' {\em
  Adv.Theor.Math.Phys.} {\bf 2} (1998) 253--291,
\href{http://www.arXiv.org/abs/hep-th/9802150}{{\tt hep-th/9802150}}.
%%CITATION = HEP-TH/9802150;%%.

\bibitem{Aharony:1999ti}
O.~Aharony, S.~S. Gubser, J.~M. Maldacena, H.~Ooguri, and Y.~Oz, ``{Large N
  field theories, string theory and gravity},'' {\em Phys.Rept.} {\bf 323}
  (2000) 183--386,
\href{http://www.arXiv.org/abs/hep-th/9905111}{{\tt hep-th/9905111}}.
%%CITATION = HEP-TH/9905111;%%.

\bibitem{Grumiller:2008qz}
D.~Grumiller and N.~Johansson, ``{Instability in cosmological topologically
  massive gravity at the chiral point},'' {\em JHEP} {\bf 0807} (2008) 134,
\href{http://www.arXiv.org/abs/0805.2610}{{\tt 0805.2610}}.
%%CITATION = ARXIV:0805.2610;%%.

\bibitem{Grumiller:2013at}
D.~Grumiller, W.~Riedler, J.~Rosseel, and T.~Zojer, ``{Holographic applications
  of logarithmic conformal field theories},'' {\em J.Phys.} {\bf A46} (2013)
  494002,
\href{http://www.arXiv.org/abs/1302.0280}{{\tt 1302.0280}}.
%%CITATION = ARXIV:1302.0280;%%.

\bibitem{Vafa:2014iua}
C.~Vafa, ``{Non-Unitary Holography},''
\href{http://www.arXiv.org/abs/1409.1603}{{\tt 1409.1603}}.
%%CITATION = ARXIV:1409.1603;%%.

\bibitem{Sezgin:2002rt}
E.~Sezgin and P.~Sundell, ``{Massless higher spins and holography},'' {\em
  Nucl.Phys.} {\bf B644} (2002) 303--370,
\href{http://www.arXiv.org/abs/hep-th/0205131}{{\tt hep-th/0205131}}.
%%CITATION = HEP-TH/0205131;%%.

\bibitem{Klebanov:2002ja}
I.~Klebanov and A.~Polyakov, ``{AdS dual of the critical O(N) vector model},''
  {\em Phys.Lett.} {\bf B550} (2002) 213--219,
\href{http://www.arXiv.org/abs/hep-th/0210114}{{\tt hep-th/0210114}}.
%%CITATION = HEP-TH/0210114;%%.

\bibitem{Henneaux:2010xg}
M.~Henneaux and S.-J. Rey, ``{Nonlinear $W_{\infty}$ as Asymptotic Symmetry of
  Three-Dimensional Higher Spin Anti-de Sitter Gravity},'' {\em JHEP} {\bf
  1012} (2010) 007,
\href{http://www.arXiv.org/abs/1008.4579}{{\tt 1008.4579}}.
%%CITATION = ARXIV:1008.4579;%%.

\bibitem{Campoleoni:2010zq}
A.~Campoleoni, S.~Fredenhagen, S.~Pfenninger, and S.~Theisen, ``{Asymptotic
  symmetries of three-dimensional gravity coupled to higher-spin fields},''
  {\em JHEP} {\bf 1011} (2010) 007,
\href{http://www.arXiv.org/abs/1008.4744}{{\tt 1008.4744}}.
%%CITATION = ARXIV:1008.4744;%%.

\bibitem{Gaberdiel:2010pz}
M.~R. Gaberdiel and R.~Gopakumar, ``{An $\text{AdS}_3$ Dual for Minimal Model
  CFTs},'' {\em Phys.Rev.} {\bf D83} (2011) 066007,
\href{http://www.arXiv.org/abs/1011.2986}{{\tt 1011.2986}}.
%%CITATION = ARXIV:1011.2986;%%.

\bibitem{Susskind:1998vk}
L.~Susskind, ``{Holography in the flat space limit},''
  \href{http://www.arXiv.org/abs/hep-th/9901079}{{\tt hep-th/9901079}}.
[AIP Conf. Proc.493,98(1999)].
%%CITATION = HEP-TH/9901079;%%.

\bibitem{Polchinski:1999ry}
J.~Polchinski, ``{S matrices from AdS space-time},''
\href{http://www.arXiv.org/abs/hep-th/9901076}{{\tt hep-th/9901076}}.
%%CITATION = HEP-TH/9901076;%%.

\bibitem{Giddings:1999jq}
S.~B. Giddings, ``{Flat space scattering and bulk locality in the AdS / CFT
  correspondence},'' {\em Phys. Rev.} {\bf D61} (2000) 106008,
\href{http://www.arXiv.org/abs/hep-th/9907129}{{\tt hep-th/9907129}}.
%%CITATION = HEP-TH/9907129;%%.

\bibitem{Gary:2009mi}
M.~Gary and S.~B. Giddings, ``{The Flat space S-matrix from the AdS/CFT
  correspondence?},'' {\em Phys. Rev.} {\bf D80} (2009) 046008,
\href{http://www.arXiv.org/abs/0904.3544}{{\tt 0904.3544}}.
%%CITATION = ARXIV:0904.3544;%%.

\bibitem{Strominger:2001pn}
A.~Strominger, ``{The dS / CFT correspondence},'' {\em JHEP} {\bf 10} (2001)
  034,
\href{http://www.arXiv.org/abs/hep-th/0106113}{{\tt hep-th/0106113}}.
%%CITATION = HEP-TH/0106113;%%.

\bibitem{Anninos:2011ui}
D.~Anninos, T.~Hartman, and A.~Strominger, ``{Higher Spin Realization of the
  dS/CFT Correspondence},''
\href{http://www.arXiv.org/abs/1108.5735}{{\tt 1108.5735}}.
%%CITATION = ARXIV:1108.5735;%%.

\bibitem{Kachru:2008yh}
S.~Kachru, X.~Liu, and M.~Mulligan, ``{Gravity Duals of Lifshitz-like Fixed
  Points},'' {\em Phys.Rev.} {\bf D78} (2008) 106005,
\href{http://www.arXiv.org/abs/0808.1725}{{\tt 0808.1725}}.
%%CITATION = ARXIV:0808.1725;%%.

\bibitem{Son:2008ye}
D.~Son, ``{Toward an AdS/cold atoms correspondence: A Geometric realization of
  the Schrodinger symmetry},'' {\em Phys.Rev.} {\bf D78} (2008) 046003,
\href{http://www.arXiv.org/abs/0804.3972}{{\tt 0804.3972}}.
%%CITATION = ARXIV:0804.3972;%%.

\bibitem{Balasubramanian:2008dm}
K.~Balasubramanian and J.~McGreevy, ``{Gravity duals for non-relativistic
  CFTs},'' {\em Phys.Rev.Lett.} {\bf 101} (2008) 061601,
\href{http://www.arXiv.org/abs/0804.4053}{{\tt 0804.4053}}.
%%CITATION = ARXIV:0804.4053;%%.

\bibitem{sachdev2011quantum}
S.~Sachdev, {\em Quantum phase transitions}.
\newblock Cambridge University Press, 2011.

\bibitem{Gary:2012ms}
M.~Gary, D.~Grumiller, and R.~Rashkov, ``{Towards non-AdS holography in
  3-dimensional higher spin gravity},'' {\em JHEP} {\bf 1203} (2012) 022,
\href{http://www.arXiv.org/abs/1201.0013}{{\tt 1201.0013}}.
%%CITATION = ARXIV:1201.0013;%%.

\bibitem{Afshar:2012hc}
H.~Afshar, M.~Gary, D.~Grumiller, R.~Rashkov, and M.~Riegler, ``{Semi-classical
  unitarity in 3-dimensional higher-spin gravity for non-principal
  embeddings},'' {\em Class. Quant. Grav.} {\bf 30} (2013) 104004,
\href{http://www.arXiv.org/abs/1211.4454}{{\tt 1211.4454}}.
%%CITATION = ARXIV:1211.4454;%%.

\bibitem{Gutperle:2013oxa}
M.~Gutperle, E.~Hijano, and J.~Samani, ``{Lifshitz black holes in higher spin
  gravity},'' {\em JHEP} {\bf 1404} (2014) 020,
\href{http://www.arXiv.org/abs/1310.0837}{{\tt 1310.0837}}.
%%CITATION = ARXIV:1310.0837;%%.

\bibitem{Afshar:2014cma}
H.~Afshar, T.~Creutzig, D.~Grumiller, Y.~Hikida, and P.~B. Ronne, ``{Unitary
  W-algebras and three-dimensional higher spin gravities with spin one
  symmetry},'' {\em JHEP} {\bf 06} (2014) 063,
\href{http://www.arXiv.org/abs/1404.0010}{{\tt 1404.0010}}.
%%CITATION = ARXIV:1404.0010;%%.

\bibitem{Gutperle:2014aja}
M.~Gutperle and Y.~Li, ``{Higher Spin Lifshitz Theory and Integrable
  Systems},'' {\em Phys. Rev.} {\bf D91} (2015), no.~4, 046012,
\href{http://www.arXiv.org/abs/1412.7085}{{\tt 1412.7085}}.
%%CITATION = ARXIV:1412.7085;%%.

\bibitem{Beccaria:2015iwa}
M.~Beccaria, M.~Gutperle, Y.~Li, and G.~Macorini, ``{Higher Spin Lifshitz
  Theories and the KdV-Hierarchy},''
\href{http://www.arXiv.org/abs/1504.06555}{{\tt 1504.06555}}.
%%CITATION = ARXIV:1504.06555;%%.

\bibitem{Lei:2015ika}
Y.~Lei and S.~F. Ross, ``{Connection versus metric description for non-AdS
  solutions in higher-spin theories},'' {\em Class. Quant. Grav.} {\bf 32}
  (2015), no.~18, 185005,
\href{http://www.arXiv.org/abs/1504.07252}{{\tt 1504.07252}}.
%%CITATION = ARXIV:1504.07252;%%.

\bibitem{Lei:2015gza}
Y.~Lei and C.~Peng, ``{Higher spin holography with Galilean symmetry in general
  dimensions},''
\href{http://www.arXiv.org/abs/1507.08293}{{\tt 1507.08293}}.
%%CITATION = ARXIV:1507.08293;%%.

\bibitem{Afshar:2012nk}
H.~Afshar, M.~Gary, D.~Grumiller, R.~Rashkov, and M.~Riegler, ``{Non-AdS
  holography in 3-dimensional higher spin gravity - General recipe and
  example},'' {\em JHEP} {\bf 1211} (2012) 099,
\href{http://www.arXiv.org/abs/1209.2860}{{\tt 1209.2860}}.
%%CITATION = ARXIV:1209.2860;%%.

\bibitem{Afshar:2013vka}
H.~Afshar, A.~Bagchi, R.~Fareghbal, D.~Grumiller, and J.~Rosseel, ``{Spin-3
  Gravity in Three-Dimensional Flat Space},'' {\em Phys.Rev.Lett.} {\bf 111}
  (2013), no.~12, 121603,
\href{http://www.arXiv.org/abs/1307.4768}{{\tt 1307.4768}}.
%%CITATION = ARXIV:1307.4768;%%.

\bibitem{Gonzalez:2013oaa}
H.~A. Gonzalez, J.~Matulich, M.~Pino, and R.~Troncoso, ``{Asymptotically flat
  spacetimes in three-dimensional higher spin gravity},'' {\em JHEP} {\bf 1309}
  (2013) 016,
\href{http://www.arXiv.org/abs/1307.5651}{{\tt 1307.5651}}.
%%CITATION = ARXIV:1307.5651;%%.

\bibitem{Gary:2014mca}
M.~Gary, D.~Grumiller, S.~Prohazka, and S.-J. Rey, ``{Lifshitz Holography with
  Isotropic Scale Invariance},'' {\em JHEP} {\bf 1408} (2014) 001,
\href{http://www.arXiv.org/abs/1406.1468}{{\tt 1406.1468}}.
%%CITATION = ARXIV:1406.1468;%%.

\bibitem{Campoleoni:2014tfa}
A.~Campoleoni and M.~Henneaux, ``{Asymptotic symmetries of three-dimensional
  higher-spin gravity: the metric approach},'' {\em JHEP} {\bf 03} (2015) 143,
\href{http://www.arXiv.org/abs/1412.6774}{{\tt 1412.6774}}.
%%CITATION = ARXIV:1412.6774;%%.

\bibitem{Detournay:2005fz}
S.~Detournay, D.~Orlando, P.~M. Petropoulos, and P.~Spindel,
  ``{Three-dimensional black holes from deformed anti-de Sitter},'' {\em JHEP}
  {\bf 07} (2005) 072,
\href{http://www.arXiv.org/abs/hep-th/0504231}{{\tt hep-th/0504231}}.
%%CITATION = HEP-TH/0504231;%%.

\bibitem{Anninos:2008fx}
D.~Anninos, W.~Li, M.~Padi, W.~Song, and A.~Strominger, ``{Warped AdS(3) Black
  Holes},'' {\em JHEP} {\bf 03} (2009) 130,
\href{http://www.arXiv.org/abs/0807.3040}{{\tt 0807.3040}}.
%%CITATION = ARXIV:0807.3040;%%.

\bibitem{Jeong:2014iva}
J.~Jeong, E.~Ó~Colgáin, and K.~Yoshida, ``{SUSY properties of warped
  $AdS_3$},'' {\em JHEP} {\bf 06} (2014) 036,
\href{http://www.arXiv.org/abs/1402.3807}{{\tt 1402.3807}}.
%%CITATION = ARXIV:1402.3807;%%.

\bibitem{Deser:1982vy}
S.~Deser, R.~Jackiw, and S.~Templeton, ``{Three-Dimensional Massive Gauge
  Theories},'' {\em Phys. Rev. Lett.} {\bf 48} (1982)
975--978.
%%CITATION = PRLTA,48,975;%%.

\bibitem{Deser:1981wh}
S.~Deser, R.~Jackiw, and S.~Templeton, ``{Topologically Massive Gauge
  Theories},'' {\em Annals Phys.} {\bf 140} (1982) 372--411.
[Annals Phys.281,409(2000)].
%%CITATION = APNYA,140,372;%%.

\bibitem{Deser:1982a}
S.~Deser, R.~Jackiw, and S.~Templeton, ``Topologically massive gauge
  theories,'' {\em Erratum-ibid.} {\bf 185} (1988) 406.

\bibitem{Clement:1994sb}
G.~Clement, ``{Particle - like solutions to topologically massive gravity},''
  {\em Class. Quant. Grav.} {\bf 11} (1994) L115--L120,
\href{http://www.arXiv.org/abs/gr-qc/9404004}{{\tt gr-qc/9404004}}.
%%CITATION = GR-QC/9404004;%%.

\bibitem{Deser:2004wd}
S.~Deser, R.~Jackiw, and S.~Y. Pi, ``{Cotton blend gravity pp waves},'' {\em
  Acta Phys. Polon.} {\bf B36} (2005) 27--34,
\href{http://www.arXiv.org/abs/gr-qc/0409011}{{\tt gr-qc/0409011}}.
%%CITATION = GR-QC/0409011;%%.

\bibitem{Gibbons:2008vi}
G.~W. Gibbons, C.~N. Pope, and E.~Sezgin, ``{The General Supersymmetric
  Solution of Topologically Massive Supergravity},'' {\em Class. Quant. Grav.}
  {\bf 25} (2008) 205005,
\href{http://www.arXiv.org/abs/0807.2613}{{\tt 0807.2613}}.
%%CITATION = ARXIV:0807.2613;%%.

\bibitem{Ertl:2010dh}
S.~Ertl, D.~Grumiller, and N.~Johansson, ``{All stationary axi-symmetric local
  solutions of topologically massive gravity},'' {\em Class. Quant. Grav.} {\bf
  27} (2010) 225021,
\href{http://www.arXiv.org/abs/1006.3309}{{\tt 1006.3309}}.
%%CITATION = ARXIV:1006.3309;%%.

\bibitem{Anninos:2010pm}
D.~Anninos, G.~Compere, S.~de~Buyl, S.~Detournay, and M.~Guica, ``{The Curious
  Case of Null Warped Space},'' {\em JHEP} {\bf 11} (2010) 119,
\href{http://www.arXiv.org/abs/1005.4072}{{\tt 1005.4072}}.
%%CITATION = ARXIV:1005.4072;%%.

\bibitem{Blencowe:1988gj}
M.~Blencowe, ``{A Consistent Interacting Massless Higher Spin Field Theory in
  $D$ = (2+1)},'' {\em Class.Quant.Grav.} {\bf 6} (1989)
443.
%%CITATION = CQGRD,6,443;%%.

\bibitem{Hoppephdthesis}
J.~Hoppe, {\em Quantum theory of a massless relativistic surface and a
  two-dimensional bound state problem}.
\newblock PhD thesis, Massachusetts Institute of Technology, 1982.

\bibitem{Bergshoeff:1989ns}
E.~Bergshoeff, M.~Blencowe, and K.~Stelle, ``{Area Preserving Diffeomorphisms
  and Higher Spin Algebra},'' {\em Commun.Math.Phys.} {\bf 128} (1990)
213.
%%CITATION = CMPHA,128,213;%%.

\bibitem{Bordemann:1989zi}
M.~Bordemann, J.~Hoppe, and P.~Schaller, ``{Infinite Dimensional Matrix
  Algebras},'' {\em Phys.Lett.} {\bf B232} (1989)
199.
%%CITATION = PHLTA,B232,199;%%.

\bibitem{Banados:1994tn}
M.~Banados, ``{Global charges in Chern-Simons field theory and the (2+1) black
  hole},'' {\em Phys.Rev.} {\bf D52} (1996) 5816,
\href{http://www.arXiv.org/abs/hep-th/9405171}{{\tt hep-th/9405171}}.
%%CITATION = HEP-TH/9405171;%%.

\bibitem{Polyakov:1989dm}
A.~M. Polyakov, ``{Gauge Transformations and Diffeomorphisms},'' {\em Int. J.
  Mod. Phys.} {\bf A5} (1990)
833.
%%CITATION = IMPAE,A5,833;%%.

\bibitem{Bershadsky:1990bg}
M.~Bershadsky, ``{Conformal field theories via Hamiltonian reduction},'' {\em
  Commun. Math. Phys.} {\bf 139} (1991)
71--82.
%%CITATION = CMPHA,139,71;%%.

\bibitem{Fujisawa:2012dk}
I.~Fujisawa and R.~Nakayama, ``{Second-Order Formalism for 3D Spin-3
  Gravity},'' {\em Class. Quant. Grav.} {\bf 30} (2013) 035003,
\href{http://www.arXiv.org/abs/1209.0894}{{\tt 1209.0894}}.
%%CITATION = ARXIV:1209.0894;%%.

\bibitem{Campoleoni:2011hg}
A.~Campoleoni, S.~Fredenhagen, and S.~Pfenninger, ``{Asymptotic W-symmetries in
  three-dimensional higher-spin gauge theories},'' {\em JHEP} {\bf 1109} (2011)
  113,
\href{http://www.arXiv.org/abs/1107.0290}{{\tt 1107.0290}}.
%%CITATION = ARXIV:1107.0290;%%.

\bibitem{Ammon:2011nk}
M.~Ammon, M.~Gutperle, P.~Kraus, and E.~Perlmutter, ``{Spacetime Geometry in
  Higher Spin Gravity},'' {\em JHEP} {\bf 1110} (2011) 053,
\href{http://www.arXiv.org/abs/1106.4788}{{\tt 1106.4788}}.
%%CITATION = ARXIV:1106.4788;%%.

\bibitem{Bunster:2014mua}
C.~Bunster, M.~Henneaux, A.~Perez, D.~Tempo, and R.~Troncoso, ``{Generalized
  Black Holes in Three-dimensional Spacetime},'' {\em JHEP} {\bf 1405} (2014)
  031,
\href{http://www.arXiv.org/abs/1404.3305}{{\tt 1404.3305}}.
%%CITATION = ARXIV:1404.3305;%%.

\bibitem{Gutperle:2011kf}
M.~Gutperle and P.~Kraus, ``{Higher Spin Black Holes},'' {\em JHEP} {\bf 1105}
  (2011) 022,
\href{http://www.arXiv.org/abs/1103.4304}{{\tt 1103.4304}}.
%%CITATION = ARXIV:1103.4304;%%.

\bibitem{deBoer:2014fra}
J.~de~Boer and J.~I. Jottar, ``{Boundary Conditions and Partition Functions in
  Higher Spin AdS$_3$/CFT$_2$},''
\href{http://www.arXiv.org/abs/1407.3844}{{\tt 1407.3844}}.
%%CITATION = ARXIV:1407.3844;%%.

\bibitem{Compere:2013gja}
G.~Compère and W.~Song, ``{$\mathcal{W}$ symmetry and integrability of higher
  spin black holes},'' {\em JHEP} {\bf 1309} (2013) 144,
\href{http://www.arXiv.org/abs/1306.0014}{{\tt 1306.0014}}.
%%CITATION = ARXIV:1306.0014;%%.

\bibitem{Compere:2013nba}
G.~Compère, J.~I. Jottar, and W.~Song, ``{Observables and Microscopic Entropy
  of Higher Spin Black Holes},'' {\em JHEP} {\bf 1311} (2013) 054,
\href{http://www.arXiv.org/abs/1308.2175}{{\tt 1308.2175}}.
%%CITATION = ARXIV:1308.2175;%%.

\bibitem{Henneaux:2013dra}
M.~Henneaux, A.~Perez, D.~Tempo, and R.~Troncoso, ``{Chemical potentials in
  three-dimensional higher spin anti-de Sitter gravity},'' {\em JHEP} {\bf
  1312} (2013) 048,
\href{http://www.arXiv.org/abs/1309.4362}{{\tt 1309.4362}}.
%%CITATION = ARXIV:1309.4362;%%.

\bibitem{Castro:2012bc}
A.~Castro, E.~Hijano, and A.~Lepage-Jutier, ``{Unitarity Bounds in AdS$_3$
  Higher Spin Gravity},'' {\em JHEP} {\bf 06} (2012) 001,
\href{http://www.arXiv.org/abs/1202.4467}{{\tt 1202.4467}}.
%%CITATION = ARXIV:1202.4467;%%.

\bibitem{Ryu:2006bv}
S.~Ryu and T.~Takayanagi, ``{Holographic derivation of entanglement entropy
  from AdS/CFT},'' {\em Phys. Rev. Lett.} {\bf 96} (2006) 181602,
\href{http://www.arXiv.org/abs/hep-th/0603001}{{\tt hep-th/0603001}}.
%%CITATION = HEP-TH/0603001;%%.

\bibitem{Ammon:2013hba}
M.~Ammon, A.~Castro, and N.~Iqbal, ``{Wilson Lines and Entanglement Entropy in
  Higher Spin Gravity},'' {\em JHEP} {\bf 10} (2013) 110,
\href{http://www.arXiv.org/abs/1306.4338}{{\tt 1306.4338}}.
%%CITATION = ARXIV:1306.4338;%%.

\bibitem{deBoer:2013vca}
J.~de~Boer and J.~I. Jottar, ``{Entanglement Entropy and Higher Spin Holography
  in AdS$_3$},'' {\em JHEP} {\bf 04} (2014) 089,
\href{http://www.arXiv.org/abs/1306.4347}{{\tt 1306.4347}}.
%%CITATION = ARXIV:1306.4347;%%.

\bibitem{Bagchi:2014iea}
A.~Bagchi, R.~Basu, D.~Grumiller, and M.~Riegler, ``{Entanglement entropy in
  Galilean conformal field theories and flat holography},'' {\em Phys. Rev.
  Lett.} {\bf 114} (2015), no.~11, 111602,
\href{http://www.arXiv.org/abs/1410.4089}{{\tt 1410.4089}}.
%%CITATION = ARXIV:1410.4089;%%.

\bibitem{Bagchi:2015wna}
A.~Bagchi, D.~Grumiller, and W.~Merbis, ``{Stress tensor correlators in
  three-dimensional gravity},''
\href{http://www.arXiv.org/abs/1507.05620}{{\tt 1507.05620}}.
%%CITATION = ARXIV:1507.05620;%%.

\end{thebibliography}\endgroup

\end{document}